\documentclass[useAMS,usenatbib,usegraphicx]{mn2e}

\newcommand{\Lone}{L_{1}}

\title[Computational Eulerian Hydrodynamics and Galilean Invariance]
   {Computational Eulerian Hydrodynamics\\and Galilean Invariance}

\author[B. Robertson et al.]
  {Brant E. Robertson,$^1$$^,$$^2$\thanks{Spitzer and KICP Fellow}\thanks{Current Address: Astronomy Department, California Institute of Technology, MC 249-17, 1200 East California Boulevard, Pasadena, CA 91125, USA}\thanks{brant@astro.caltech.edu}
   Andrey V. Kravtsov,$^1$$^,$$^2$
   Nickolay Y. Gnedin,$^1$$^,$$^3$
   \newauthor
   Tom Abel$^4$
   and Douglas H. Rudd$^5$\\
   $^1$Kavli Institute for Cosmological Physics, and Department of Astronomy and
Astrophysics,\\University of Chicago, 933 East 56th Street, Chicago, IL 60637, USA \\
   $^2$Enrico Fermi Institute, 5640 South Ellis Avenue, Chicago, IL 60637, USA\\
   $^3$Particle Astrophysics Center, Fermilab, Batavia, IL 60510, USA\\
   $^4$Kavli Institute for Particle Astrophysics and Cosmology, Stanford University,\\2575 Sand Hill Road, Menlo Park, CA 94025, USA\\
   $^5$School of Natural Sciences, Institute for Advanced Study, Princeton, NJ 08540, USA}
\date{Released 2009 Xxxxx XX}

\pagerange{\pageref{firstpage}--\pageref{lastpage}} \pubyear{2009}

\begin{document}

\label{firstpage}

\maketitle

\begin{abstract}
Eulerian hydrodynamical simulations are a powerful and popular tool
for modeling fluids in astrophysical systems. 
In this work, we critically examine recent claims that these
methods violate Galilean invariance of the Euler equations. 
We demonstrate that Eulerian hydrodynamics methods do converge to
a Galilean-invariant solution, provided a well-defined
convergent solution exists. Specifically, we show that
numerical diffusion, resulting from diffusion-like terms in the
discretized hydrodynamical equations solved by Eulerian methods,
accounts for the effects previously identified as evidence for the
Galilean non-invariance of these methods.  These velocity-dependent
diffusive terms lead to different results for different bulk velocities {\it
when the spatial resolution of the simulation is kept fixed,} but
their effect becomes negligible as the resolution of the simulation is
increased to obtain a converged solution.  In particular, we find that
Kelvin-Helmholtz instabilities develop properly in realistic Eulerian
calculations regardless of the bulk velocity provided the problem is
simulated with sufficient resolution (a factor of 2-4 increase
compared to the case without bulk flows for realistic velocities).
Our results reiterate that high-resolution Eulerian methods can perform well and 
obtain a convergent solution, even in the presence of highly
supersonic bulk flows.
\end{abstract}

\begin{keywords}
hydrodynamics--instabilities--methods:numerical
\end{keywords}

\section{Introduction}
\label{section:introduction}

Eulerian methods have been the tool of choice in computational fluid
dynamics for over five decades.  Many successful Eulerian methods in
popular use descended from the \cite{godunov1959a} scheme that
combines the analytical Riemann solution of the Euler
equations\footnote{There are also Eulerian astrophysical hydrodynamics
codes that do not use a Godunov scheme, such as the ZEUS code
\citep{stone1992a,stone1992b,clark1996a,hayes2006a} and the code by
\cite{ryu1993a} based on the total variation diminishing flux-corrected method by
\cite{harten1983a}} with the upwind scheme of \cite{courant1952a} to
numerically evolve fluid systems on a discretized mesh.  These
Godunov-type schemes, as such methods are commonly called, have been
further engineered to include higher-order spatial reconstructions of
the fluid distribution based on piecewise linear
\citep[e.g.,][]{van_leer1977a}, parabolic \citep[e.g., PPM,
][]{colella1984a}, or, more generally, higher-order weighted
essentially non-oscillatory interpolation schemes \citep{liu1994a} .
Eulerian methods have also become quite popular for addressing
problems in Newtonian astrophysics
\citep[e.g.,][]{fryxell_etal89,cen_etal90,bryan_etal94,quilis_etal96,yepes_etal97,wada_norman99,ricker_etal00},
especially in the framework of Adaptive Mesh Refinement \citep[AMR,
e.g.,][]{bryan1997a,khokhlov98,truelove_etal98,fryxell2000a,plewa_mueller01,kravtsov_etal02,teyssier2002a,quilis04,wang2008a}.
Given their wide-spread use in computational astrophysics, an
understanding of the fundamental limitations of such codes is
important for interpreting the astrophysics of hydrodynamical systems
that cannot be accessed through laboratory experiments.

While Eulerian astrophysical simulation codes routinely demonstrate
excellent performance on idealized test cases, some shortcomings of
these methods are known \citep[e.g.,][]{quirk94,quirk05}. 
Recently, several
studies have focused on the differences produced by Eulerian codes
in reference frames moving with different velocities with respect
to the computational grid. 
\cite{wadsley2008a}
emphasized the role of diffusion in altering the development of
Kelvin-Helmholtz instabilities in the FLASH code \citep{fryxell2000a}
simulations of bouyant, hot bubbles.  \cite{tasker2008a} simulated the
advection of otherwise static, self-gravitating gas clouds, and showed
that the performance of FLASH and the PPM version of Enzo
\citep{bryan1997a,bryan1999a,norman1999a,bryan2001a,oshea2004a} 
in maintaining the centroid and density profile of
the gas cloud depended on its velocity with 
respect to the static computational grid.  Most recently,
\cite{springel2009a} motivated the development of the new 
Lagrangian-Eulerian moving-mesh code AREPO by demonstrating that with
fixed grid Godunov solvers Kelvin-Helmholtz instabilities may not
develop and evolve properly when the interface between the two fluids
has a large bulk velocity with respect to the grid.
These apparent
failures of Eulerian codes have been discussed in terms of ``Galilean
non-invariance,'' which in this context means that for initial
conditions that move with different uniform bulk velocities with
respect to the computational grid but are otherwise identical,
numerical solutions obtained with Eulerian codes may depend on the
chosen bulk velocity.

Given the ubiquity of supersonic bulk motions in astrophysical
scenarios, these results are potentially damning for the application
of stationary mesh Eulerian codes to galaxy and structure formation.
The purpose of this work is to critically examine the performance of
Eulerian hydrodynamical codes for simulating systems with supersonic
bulk motions, and to clarify both the nature and meaning of the
velocity-dependent differences highlighted in previous studies.
Specifically, we use the Eulerian mesh codes ART \citep{kravtsov_etal02}
and Enzo 
\citep{bryan1997a,bryan1999a,norman1999a,bryan2001a,oshea2004a} 
to simulate the
development of Kelvin-Helmholtz instabilities in test calculations
similar to those presented in \cite{springel2009a}. We employ
statistical measures to quantify convergence and error of the
calculations in addition to an extensive visual comparison of the
solutions. We show that the effects discussed above are not a
consequence of Galilean non-invariance of Riemann solvers, but rather
a result of diffusive errors accumulated during advection of fluid
through the computational grid.  The effects of these errors are thus
particularly acute in systems where perturbations and the interface
between fluids are under-resolved. We demonstrate that with a proper
initial setup the Eulerian methods produce a convergent solution at
large bulk velocity as the resolution of the simulation is increased.

The paper is organized as follows.  In \S \ref{section:diffusion}, we discuss the 
origin of numerical diffusion in the Eulerian method and illustrate
its effects using simulations of contact discontinuities.  
Readers familiar with the effects of numerical diffusion should
proceed to \S \ref{section:kh}, where we review
previous simulations of Kelvin-Helmholtz instabilities and the related
claims of Galilean non-invariance in Eulerian methods.  In \S
\ref{section:kh_abel}, we present a new, better-behaved test
calculation of Kelvin-Helmholtz instability and study the development
of the instability over a range of resolutions and supersonic bulk
motions.  We study the statistical and error properties of the
Kelvin-Helmholtz simulations and use these statistics to critically
examine the apparent Galilean non-invariance of Eulerian simulation
codes.  We discuss our results in \S \ref{section:discussion} and
present a summary in \S \ref{section:summary}.

\section{Numerical Diffusion}
\label{section:diffusion}

Computational Eulerian hydrodynamical codes calculate the evolution
of fluid systems using a discretized approximation to Euler's
equations.  When modeling the conservative form of Euler's equations, 
the change in 
quantities like density or energy integrated over cell units of size
$\Delta x$
in the discretized mesh over a time step $\Delta t$ will correspond
to the flux of those quantities across the cell boundaries over the
same time interval.  Fluid interactions between cells then fundamentally
involve calculations of the fluxes, which can be approximated using 
solutions to the Riemann problem \citep[i.e.,][]{godunov1959a} or
through other means 
\citep[e.g., the flux corrected methods of][and \citealt{harten1983a}, see also Chapter 21 of \citealt{laney1998a}]{boris1973a}.
Since these numerical approximations to the physical fluxes exchanged
between fluid volumes during the time interval $\Delta t$ 
are discretized, there is a truncation error associated with the
numerical approximation.  Missing or extraneous higher order terms in the
discretized numerical approximation can appear as an effective viscosity or
thermal conductivity and lead to the smearing or
dispersion of features in fluid flow.  We will refer to smearing
effects as numerical diffusion,
while effects that change the wave
speed of features in the fluid will be labeled numerical dispersion
\footnote{We note that the numerical diffusion owing to truncation error in the
Eulerian method is very distinct from artificial viscosity employed in
Smoothed Particle Hydrodynamics to improve shock capturing, and the two should not be confused.}.
Clear discussions about the effects of numerical diffusion can be found
in \cite{boris1973a} and \cite{laney1998a}.

The strength of numerical diffusion depends on the method chosen
to model fluid systems.  Lagrangian methods integrate the convective
derivative form of the mass conservation equation directly, and therefore suffer 
from small diffusive truncation errors. 
Eulerian methods calculate the advective term in the mass conservation
equation explicitly, which can lead to an appreciable diffusive truncation error upon
discretization.  Some Eulerian methods, such as Flux-Corrected Transport
algorithms \cite[e.g.,][]{boris1973a}, include an explicit numerical diffusion
term proportional to a second spatial derivative that owes to their forced
conservative and non-negative properties (the ``flux correction'' refers to the explicit
artificial anti-diffusion used to correct this truncation error term).

For Godunov-type methods based on Riemann solvers, differences
in the amount of numerical diffusion can arise from the approximations made
in constructing the
discrete representation of the local fluid flow on the computational
mesh.  In Godunov-type methods, the numerical flux between cells is
determined by the known solution of the piecewise-constant Riemann problem.  
The resulting flux across the cell face is then determined by the properties
of fluid states on either side of the face, the cell size, and the time step size.
Resolution determines the region used to average the 
fluid properties for finding the initial states in the Riemann problem.  The
averaging procedure introduces numerical diffusion, and can be counter-acted
by higher spatial resolution.  Improving the quality of the approximation to the
fluid states used in the Riemann problem can also decrease the amount of
numerical diffusion, so the method used to model the shape of the fluid flow
on the grid can change the diffusivity of the method.  For instance, the 
local flow 
can be approximated by constant \citep{godunov1959a}, linear \citep{van_leer1977a}, 
parabolic \citep{colella1984a}, or higher-order piecewise
polynomial \citep{liu1994a} interpolations on the discrete mesh.  Higher-order interpolations
improve the local approximations used in reconstructing the fluid
flow and calculating the
initial states to the Riemann problem, and therefore will suffer from less numerical diffusion.
In general, the strength of numerical diffusion will also depend on the local flow velocity.
This velocity dependence arises because, in the presence of a large advective flow, more time
steps are used and more local averages are performed.  Additionally, with a large bulk velocity a 
larger (Lagrangian) region of the fluid is averaged to calculate the 
input states for the Riemann problem.

We can illustrate how numerical diffusion affects the shape of the local
fluid distribution by simulating the advection of contact discontinuities.
In the absence of numerical diffusion, the square wave should be perfectly advected
and the contact discontinuities would remain sharp.  
However, as these simple tests
will illustrate, numerical diffusion will act to soften the contact discontinuities
in a resolution- and velocity-dependent manner.  The effects of numerical diffusion in 
these tests will prove to be informative for simulations of the Kelvin-Helmholtz instability
in \S \ref{section:kh} and \ref{section:kh_abel}. 

For the simple problem of the advection of a waveform with a constant
velocity $v$, the advected quantity $\rho$ (e.g., the density)
obeys the partial differential advection equation
\begin{equation}
\label{equation:advection}
\frac{\partial \rho}{\partial t} + v \frac{\partial \rho}{\partial x} = 0,
\end{equation}
\noindent
as a function of position $x$ and time $t$.  The solution of this
equation is simply $\rho(x,t) = \rho(x-vt,0)$, as the initial
waveform advects with a constant velocity $v$.  

However, as discussed
by \citet[][see, e.g., his \S 5.2.1]{toro1997a}, numerical methods for solving
the advection equation (or Euler's equations) actually solve a slightly 
{\it modified} equation (to some approximate order).  For example, in the case of
the simple first-order upwind scheme of \cite{courant1952a} the modified
equation solved by the numerical method is
\begin{equation}
\label{equation:advection_modified}
\frac{\partial \rho}{\partial t} + v \frac{\partial \rho}{\partial x} = \alpha \frac{\partial^{2} \rho}{\partial x^{2}}.
\end{equation}
\noindent
In this advection-diffusion equation the right-hand side 
acts as a form of numerical diffusion with a diffusion constant $\alpha$.  The
solution of Equation \ref{equation:advection_modified} will differ from the
solution of Equation \ref{equation:advection} if $\alpha\ne0$, and will be
characterized by a progressive smearing of the original waveform.  The
detailed behavior of the solution to Equation \ref{equation:advection_modified}
will then depend on the diffusion constant $\alpha$.

For the first-order upwind scheme of \cite{courant1952a}, the diffusion
constant is
\begin{equation}
\label{equation:diffusion_constant}
\alpha = \frac{1}{2} v \Delta x  (1 - |c|),
\end{equation}
\noindent
where $c = v \Delta t / \Delta x$ is the Courant number (numerical stability 
requires $|c|\leq1$),
$\Delta t$ is the timestep, and $\Delta x$ is the spatial grid size.
One then expects that the diffusive error induced through the 
truncation of Equation \ref{equation:advection} into Equation 
\ref{equation:advection_modified} by the discretization of the numerical scheme will
decrease with the grid size but increase with the advective velocity.
As $\Delta x\to0$ or $v\to0$, the pure advection equation (\ref{equation:advection})
is recovered.
Higher-order Eulerian methods (such as those used in this paper) 
can change the form of Equations 
\ref{equation:advection_modified} or \ref{equation:diffusion_constant},
and also introduce dispersive terms that scale as high-order odd-power
spatial derivatives.  However, as we will show, in higher-order methods
the strength of numerical
diffusion will still increase with advection velocity and decrease with 
increasing spatial resolution.  In the remainder of this section, we will use
square wave advection simulations to illustrate these numerical features of
Eulerian methods.

Unless otherwise noted,
the simulations presented in this paper use the Eulerian code ART with
piecewise-linear reconstruction and an exact Riemann solver
\citep{colella1985a}, based on the adaptive refinement strategy
developed by \cite{khokhlov98}.
For
the following square wave advection simulations, time steps were determined using the
Courant-Friedrichs-Lewy condition with a parameter $cfl=0.6$. 
ART uses a
dual-energy formulation similar to that of \cite{bryan_etal95}, such
that the internal energy equation is followed separately when the
local flow is kinetic-energy dominated and effectively pressureless.
We have checked that similar results can be obtained using an
entropy equation instead of the internal energy in the dual-energy
formulation, as discussed by \cite{ryu1993a} and \cite{springel2009a}.

\begin{figure}
\includegraphics[width=84mm]{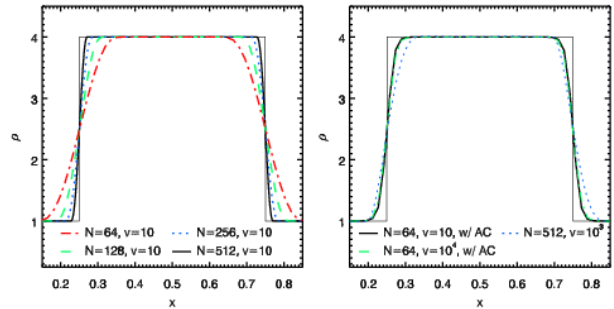}
\caption{
Simulations of a square wave contact discontinuity advected with a constant velocity.
The contact discontinuity is initialized with a density $\rho=4$ for positions $|x-0.5|\leq0.25$
and $\rho=1$ for $|x-0.5|>0.25$, and a constant pressure $P=1$ (thin black line).  The
left panel shows the contact discontinuity advected with velocity $v=10$ simulated with
resolutions $N=64$ (red dash-dotted line), $N=128$ (green dashed line), $n=256$ (blue dotted
line), and $N=512$ (thick black line) after time $t=1$.  The numerical diffusive error increases with decreasing
resolution, and tends to smear out the contact discontinuities.  The right panel shows the
same square wave advected for a time $t=1$ using a resolution of $N=64$ with advective velocities of $v=10$ (thick
black line) and $v=10^{4}$ (dashed green line), but including artificial compression in the
form of slope steepeners \citep{yang1990a,balsara1998a}.  A phase error of $8.5\%$ in the
$N=64$, $v=10^{4}$ has been corrected.  Artificial compression limits makes numerical diffusive
error roughly independent of velocity, even for advective flows with Mach number $M\approx15,000$.
Also shown is the $v=10^{4}$ simulation with $N=512$ (blue dotted line), which has been completely
smeared away by diffusive error.
}
\label{fig:cd}
\end{figure}

The
one-dimensional square wave density is initialized with $\rho=5$ for positions 
$|x-0.5|\leq0.25$ and $\rho=1$ for $|x-0.5|>0.25$.  The system
has a constant pressure $P=1$ and an adiabatic index $\gamma=5/3$.
In a first set of tests, the wave is advected to the right with a velocity $v=10$ in a periodic box 
such
that the wave travels through the box ten times over the final simulation time $t=1$.
To illustrate the role of resolution on the strength of numerical diffusion, the 
simulation is performed  with grid resolutions of $N=[64,128,256,512]$.
The left panel of Figure \ref{fig:cd} shows the final square wave density distribution
at time $t=1$ as a function of resolution, compared with the initial distribution (thin
solid line).
At low resolution ($N=64$, red dashed dotted line), numerical diffusion smears out each contact discontinuity
over approximately twelve cells, or roughly $\sim20\%$ of the computational volume.  As the resolution increases,
the contact diffuses out over more cells ($\sim18$ cells for $N=512$, black solid line) but less of the
computational volume ($\sim4\%$ for $N=512$).  The contact is physically better resolved with increasing
grid size and the diffusive error reduced.  The right panel illustrates the additional error induced by
increasing the velocity by a factor of $100$ (for $N=512$, blue dotted line).  The highest-resolution 
simulation has an increased error that degrades the effective resolution of the simulation by a factor $\sim 4$
(comparable to the $N=128$ simulation with $v=10$ shown in the left panel).  

In these one dimensional
simulations, the diffusive error can be mitigated through the use of an artificial compression (AC) technique.  Similar
in spirit to explicit anti-diffusion terms added to the flux corrected methods, AC
simply increases the slopes used in the reconstruction of the fluid on the mesh near contact discontinuities.
This approach reduces the effective second order truncation error by limiting the influence of the outer cells 
in the computational stencil.  We use the slope steepening approach of \cite{yang1990a}, as 
implemented by \cite{balsara1998a} for linear reconstruction, and refer the reader to \S 2.2 of \cite{balsara1998a} for details 
\citep[see also][for an implementation of slope steepeners for PPM]{fryxell2000a}.  The right panel of
Figure \ref{fig:cd} shows the results for a $N=64$ grid with bulk velocity $v=10$ and AC (solid
black line).  With AC the $N=64$, $v=10$ results improve to be comparable to the $N=128$, $v=10$
results without AC.  If the same $N=64$ simulation with AC is performed but with the bulk velocity
increased to $v=10^{4}$ (or Mach $M\approx15,000$, green dashed line), the results are striking.  Remarkably, with 
artificial compression the diffusive error becomes almost independent of the bulk velocity and the
$N=64$, $v=10^{4}$ simulation has almost the same diffusive error as the $N=64$, $v=10$ simulation 
(and is superior to the $N=512$, $v=100$ simulation at $10\times$ smaller bulk velocity), even as the
$N=64$, $v=10^{4}$ simulation has traversed the computational volume $10^{4}$ times and executed
$1.2\times10^{6}$ individual timesteps.  
However, a dispersive error has been introduced that changes the square wave period by $8\%$ and that
we have removed in Figure \ref{fig:cd}.  While this fractional dispersive error is only $0.08/10^{4}\sim10^{-5}$, the
error grows to an appreciable fraction of the period by $t=1$ (as discussed by \citealt{boris1973a}, this dispersive
error may also depend on the frequency of the wave form).

These simulations demonstrate the salient effects of numerical diffusion on the
properties of fluid distributions simulated with Eulerian codes.  Diffusion limits
the sharpness of fluid distributions, and the averaging of
fluid properties within cells does not preserve local discontinuities.  The effects
of diffusion can be mitigated through the use of higher spatial resolution to improve
the local reconstruction of the fluid distribution, or through intrinsically less-diffusive
methods.  The presence of a bulk advective velocity in the fluid also increases numerical diffusion
by increasing the number of time steps and local averages of the fluid distribution, and
can degrade the effective resolution of the computational grid.  However, the simulation
results naturally improve with increasing grid resolution.  With these diffusive properties of
Eulerian simulations in mind, we will now examine simulations of the development of fluid instabilities in
shearing flows.

\section{The Kelvin-Helmholtz Instability}
\label{section:kh}

\subsection{Kelvin-Helmholtz Instability with a Sharp Interface}
\label{section:kh:springel2009}

The Kelvin-Helmholtz (KH) instability \citep[][see especially Chapter XI of \citealt{chandra1961a}]{helmholtz1868a,kelvin1910a} is 
the unstable growth of perturbations at the interface between two fluid flows 
driven by shearing motions. 
Perturbations between these fluid phases that grow and become unstable
typically form waves that crest owing to the shearing motion in the
fluid.
The kinetic energy of the shearing motion powers the instability, and 
larger shear velocity gradients typically increase the proclivity for
instabilities to develop.  In the absence of viscosity and gravity, only inertia
can exert a stabilizing influence on perturbations and damp oscillations before growth
commences.

Numerical simulations of the KH instability previously studied in
astrophysical contexts include the stability of interstellar clouds in
a shearing flow \citep{murray1993a,vietri1997a,agertz2007a}, the
stripping of gas from galaxies by an intercluster medium
\citep[][]{quilis_etal00,mori2000a}, the formation and ionization
state of the Magellanic Stream \citep{bland-hawthorn2007a}, and the
survivability of high-velocity clouds \citep{heitsch2009a}.  Here, we
focus on numerical experiments of the KH instability in an idealized
setting for testing the performance of static mesh Eulerian codes in
the presence of bulk flows, but our conclusions will weigh on the
validity of the results of many such astrophysical studies.

A common choice for the initial inhomogeneity that gives rise
to the KH instability is two uniform fluids separated by a surface where the density and shearing velocities
change discontinuously.  The KH instability arising from perturbations about these initial conditions is studied 
in detail by \cite{kelvin1910a} and \citet[][\S 100]{chandra1961a}.  For a surface discontinuity, 
the growth of any perturbations about the surface can be calculated from 
the Euler equations by separating the solution into normal modes.  As discussed by \cite{chandra1961a}, 
instability will 
occur 
at a perfectly discontinuous interface 
regardless of the magnitude of the shearing velocity.  For such
initial conditions, this
result holds generally for some minimum wavenumber, and in the
absence of gravity or surface tension applies to {\it all wavenumbers.\/}

\begin{figure}
\includegraphics[width=84mm]{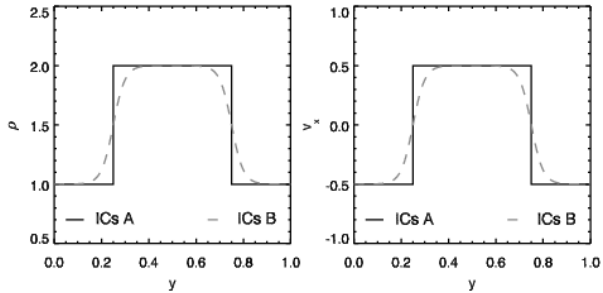}
\caption{Kelvin-Helmholtz (KH) instability simulation initial conditions for the density (left panel) and
$x$-direction shear velocity (right panel) as a function of $y$-position.
Shown are the initial conditions for the Springel (2009) KH simulations (``ICs A'', black line),
as well as a new KH simulation with a smoothly-varying density and shear velocity (``ICs B'', dashed gray line).
Both simulations have additional $y$-direction velocity perturbations to seed the instability (see text).
The system has a constant pressure $P=2.5$ and adiabatic index $\gamma=5/3$.}
\label{fig:kh_ics}
\end{figure}

The discontinuous, two-fluid KH instability has been used as a test simulation in recent
years for evaluating the performance of hydrodynamical codes.
\cite{agertz2007a} studied the relative performance of smoothed particle hydrodynamics and Eulerian
grid codes in calculating the development of KH instabilities from two-fluid initial conditions.
\cite{springel2009a} also studied KH instabilities in shearing, sharp interface between two 
fluids in two-dimensional simulations to test the performance
of the Eulerian scheme in the presence of bulk flows.  The discontinuous, 
two-fluid KH instability simulations of \cite{springel2009a}
were performed in a unit computational volume in the $x-y$ plane with periodic boundaries.
The initial conditions consisted of a central fluid slab at $|y-0.5|<0.25$ with density $\rho_{1}=2$ and $v_{1}=0.5$ (Mach $M=0.35$) 
surrounded by a second fluid at $|y-0.5|>0.25$ with density $\rho_{2}=1$ and $v_{2}=-0.5$ (Mach $M=0.25$).  The fluids
were initialized in pressure equilibrium with $P=2.5$ and an adiabatic index of $\gamma=5/3$.  A
sinusoidal velocity perturbation in the $y$-direction of the form 
\begin{eqnarray}
\label{equation:volker_perturbation}
v_{y}(x,y) &=& w_{0}\sin(n\pi x) \times\\
&&\left\{\exp\left[-\frac{(y-0.25)^2}{2\sigma^{2}}+\frac{(y-0.75)^2}{2\sigma^{2}}\right]\right\}\nonumber,
\end{eqnarray}
\noindent
with parameters $n=4$, $w_{0}=0.1$, and $\sigma=0.05$
was added to provide a seed for the instability.  For reference, Figure \ref{fig:kh_ics} shows the 
density and shearing velocity initial conditions.  We will refer to these discontinuous, two-fluid KH 
instability initial conditions as ``ICs A''.   When discussing the Mach number of any bulk motions, we
will refer to the Mach number relative to the sound speed in the dense fluid unless otherwise noted.

\cite{springel2009a} evolved the system for a time $t=2$ using the new
moving-mesh code AREPO using an exact Riemann solver \citep{toro1997a}
in a fixed-mesh mode, and with the Eulerian PPM code Athena
\citep{stone2008a} using the linearized solver of \cite{roe1984a}.
The KH simulations presented in this paper use the Eulerian code ART,
with the method described in \S \ref{section:diffusion}, unless otherwise noted. 
As is customary, the ART code uses
\cite{strang1968a} dimensional splitting to numerically integrate the
multidimensional Euler equations, but we have checked that using the
unsplit solver of \cite{gardiner2008a} produces similar results.  
For
the presented KH simulations, time steps were determined using the
Courant-Friedrichs-Lewy condition with a parameter $cfl=0.6$, except
near snapshot times where time steps were determined by requiring a
simulation output every $\Delta t=0.01$ time interval.  

\begin{figure*}
\includegraphics[width=168mm]{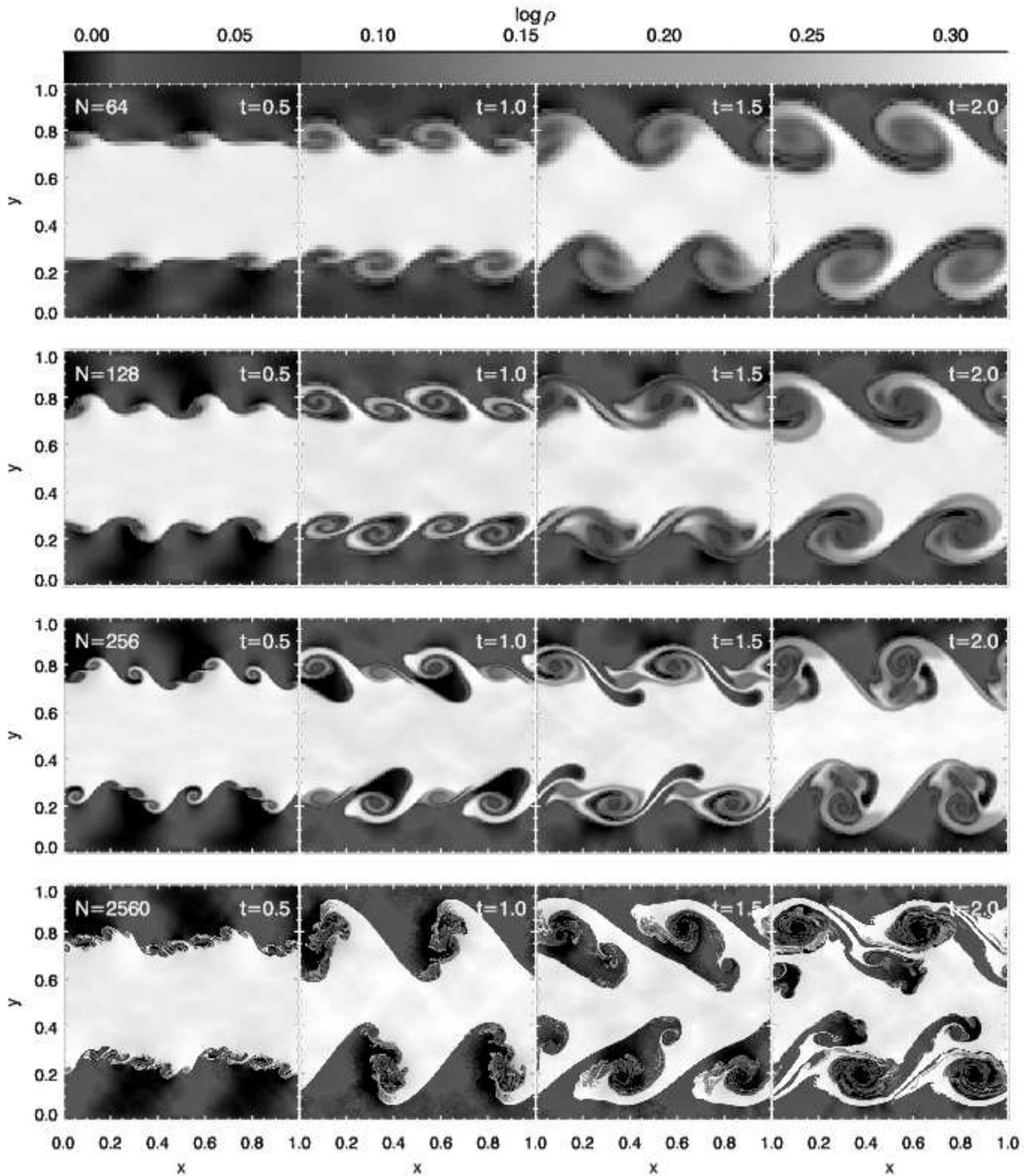}
\caption{
Kelvin-Helmholtz instability simulation of ICs A.  Shown is the temporal evolution of
the simulation with a mesh resolution of
$64\times64$ (first row), $128\times128$ (second row),  $256\times256$ (third row),
and $2560\times2560$ (fourth row) at times $t=0.5$ (first column), $t=1.0$ (second column), 
$t=1.5$ (third column), and $t=2.0$ (fourth column).
}
\label{fig:kh_sequence}
\end{figure*} 

Figure \ref{fig:kh_sequence} shows the temporal evolution of this
simulation calculated on a fixed mesh of size
$64\times64$ (first row), $128\times128$ (second row),  $256\times256$ (third row),
and $2560\times2560$ (fourth row) at times $t=0.5$ (first column), $t=1.0$ (second column), 
$t=1.5$ (third column), and $t=2.0$ (fourth column).  A KH instability
develops in each simulation, but the detailed structure of the
growing instability differs between simulations with different grid size.  The dominant
structure in the KH instability is the $n=4$ mode seeded by the initial perturbations 
(following Equation \ref{equation:volker_perturbation}).
However, a secondary set of small-scale eddies
that {\it have not been seeded} in the initial conditions also develop. 
The development of these small-scale instabilities that increase in complexity with increasing
numerical resolution can be directly related to the chosen fluid interface.  As
noted by \cite{chandra1961a}, the discontinuous density and shearing velocity distributions
of the initial conditions allow for perturbations of all wavenumbers to be unstable to growth.
An increase in the resolution broadens the range of unstable wavelengths available for
excitation by, e.g., secondary waves generated by the seeded $n=4$ instability or numerical noise,
and we further discuss these mechanisms below.

\begin{figure*}
\includegraphics[width=168mm]{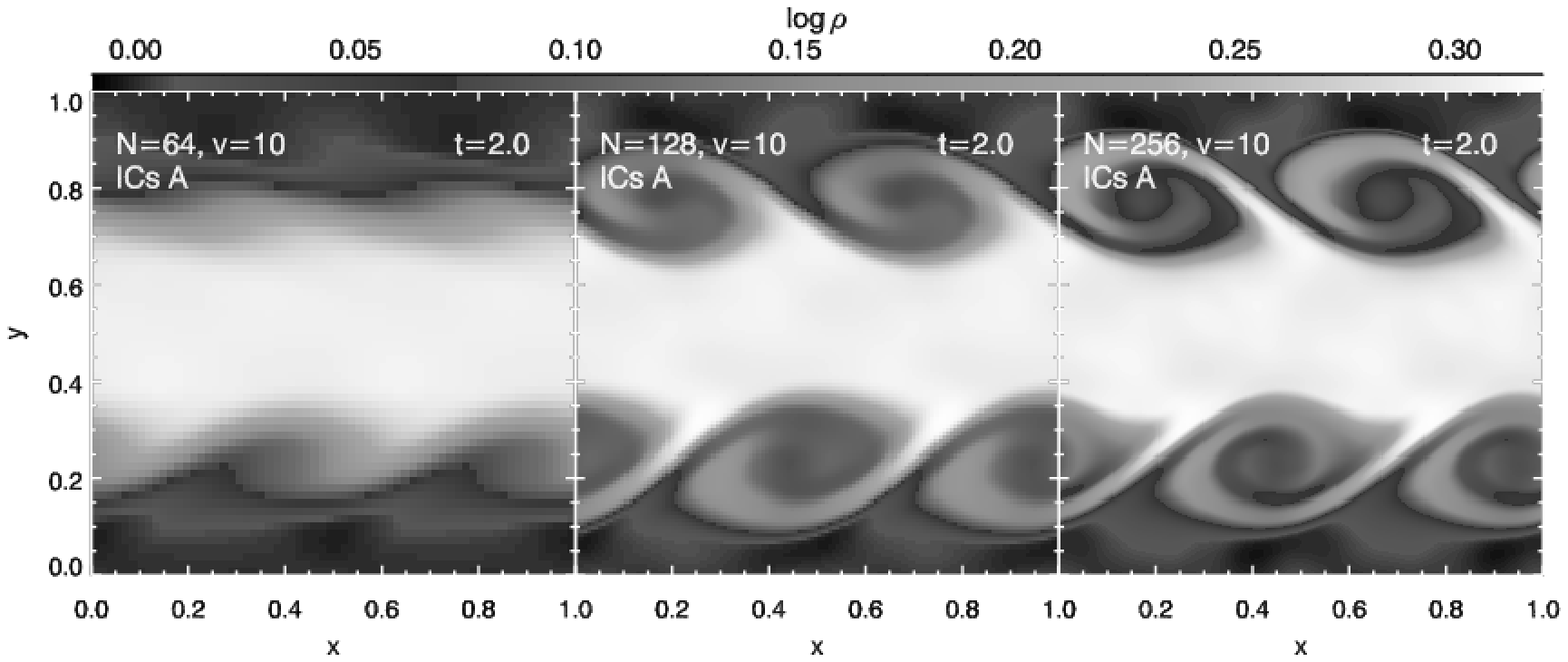}
\caption{
Kelvin-Helmholtz instability simulation of ICs A including a uniform bulk flow of $v=10$ (Mach $M=6.9$)
in the $y$-direction.  Shown is the computational grid at time $t=2$, corresponding to
ten full advections of the fluid through the box.  The simulation was performed on a 
mesh with $N=64$ (left panel), $N=128$ (middle panel), and $N=256$ (right panel) cells
on a side.  These results can be compared directly with the simulation results at $t=2$
shown in Figure \ref{fig:kh_sequence}.  Note that the instability fails to develop with
resolution $N=64$.
}
\label{fig:kh_vel_sequence}
\end{figure*}

As demonstrated by \cite{springel2009a}, the simulation of the initial conditions ICs A 
changes dramatically if a uniform bulk flow is added to the fluid.  
Figure \ref{fig:kh_vel_sequence} shows the results of the simulation at $t=2$ of ICs A
with a bulk flow of $v=10$ (Mach $M=6.9$) in the $y$-direction with resolutions of $N=64$ (left panel),
$N=128$ (middle panel), and $N=256$.  Each of these panels can be compared directly with
the results at $t=2$ for the same resolution in Figure \ref{fig:kh_sequence}
and are clearly quite different.
\cite{springel2009a} states that these differences are ``direct evidence for a violation of
Galilean invariance of the Eulerian approach.''  
Although our results clearly confirm that the KH instability
does not develop in the $N=64$ simulation, Figure
\ref{fig:kh_vel_sequence} shows that the instability does develop at
higher resolution and hints at a convergence toward a single prominent
$n=4$ mode instability.

While qualitative differences between the results of the KH instability simulation using
initial conditions ICs A are apparent in Figures \ref{fig:kh_sequence} and 
\ref{fig:kh_vel_sequence}, a quantitative comparison would be preferable.  A common
characterization of a simulation with a known solution is the error norm, such as
the $\Lone$ error norm given by
\begin{equation}
\label{equation:l1_error}
\Lone = \frac{1}{N}\sum_{i=1}^{N} |f_{i} - f_{\mathrm{true}}|,
\end{equation}
\noindent
where $N$ is the number of computational cells, $f_{i}$ is a property of
the $i$th cell, and $f_{\mathrm{true}}$ is the ``true'' property of same
cell in the known solution, and the summation runs over all $N$ cells.  
Unfortunately, error norms are useless for 
evaluating the KH simulation of ICs A because there is no convergence with
increasing $N$ and no known solution.  However, other useful statistical 
measures can be constructed to provide a quantitative gauge of the qualitative
differences.  For instance, the global correlation of the simulations
at fixed time could be compared using, e.g., Pearson's product-moment coefficient.
However, the instabilities develop over a limited range of the computational volume
and simulations of ICs A with qualitatively very different development of the
instability would be highly correlated.  A more useful, targeted statistic would track
the growth and amount of mixing in the instability, but discriminate between the sharp
features present in the simulations of Figure \ref{fig:kh_sequence} and the diffusive
features present in Figure \ref{fig:kh_vel_sequence}.  

After some experimentation, a unit-free measure of the root-mean-squared fluctuations
in the simulation at fixed $y$-position was found to provide a useful description of
instability growth and complexity.  For a property $f$, the average $\langle f\rangle$ and
variance $\sigma_{f}^{2}$ for each row in the computational mesh is calculated.
The ratio $\sigma_{f}/\langle f\rangle$ is then averaged over the computational volume
as
\begin{equation}
\label{equation:mixing_statistic}
\Sigma \sigma_{f}/\langle f\rangle = \left[ \int \mathrm{d}y \sigma_{f}(y)/\langle f(y)\rangle\right] / \left[ \int \mathrm{d}y \right].
\end{equation}
\noindent
We will refer to the quantity defined by Equation \ref{equation:mixing_statistic} as
the ``mixing statistic''.  Analogous mixing measures were used by \cite{wadsley2008a}.
Throughout the rest of the paper, when the
$\Lone$ error norm or mixing statistic are used to compare simulations of differing resolutions, the
simulations are rebinned to the minimum resolution (usually $N=64$) using the IDL function 
``CONGRID'' with the cubic interpolation value set to $\mathrm{CUBIC}=-0.5$ \citep{park1983a}.
Simulations with bulk flows are shifted to align with the computational grid as if the measurements
were performed in the moving frame.

\begin{figure}
\includegraphics[width=84mm]{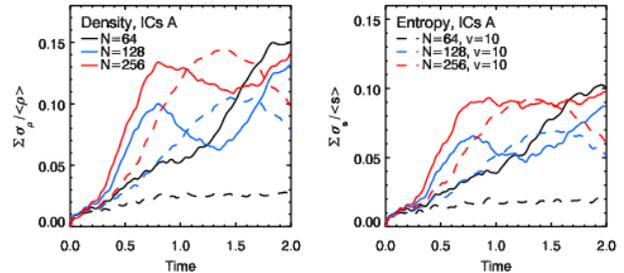}
\caption{
``Mixing statistic'' for the Kelvin-Helmholtz instability simulation of 
ICs A with time (see Equation \ref{equation:mixing_statistic}). 
Shown is a dimensionless measure of the root-mean-squared
density (left panel) and entropy (right panel) fluctuations from the growing
KH instabilities in simulations with $N=64$ (black), $N=128$ (blue), and
$N=256$ (red) resolution and without bulk flows, as well as $N=64$ (black dashed),
$N=128$ (blue dashed), and $N=256$ (red dashed) simulations with $v=10$ (Mach $M=6.9$) velocity bulk flow along the
$y$-direction.  In the simulations without bulk flows, the instabilities
grow at different rates and by different amounts.  In the simulations with
bulk flows, the instabilities become better defined with increasing resolution
but have yet to converge at $N=256$ resolution.  The instability mostly fails
to develop for the lowest resolution ($N=64$) simulation with $v=10$ bulk 
flows, as noted by Springel (2009).
}
\label{fig:kh_statistics}
\end{figure}

Figure \ref{fig:kh_statistics}
shows the mixing statistic for the KH instability simulation of ICs A as a function of
time.  Shown are the mixing statistics for the density $\rho$ and the entropy function 
$s=P/\rho^{\gamma}$ for simulations with $N=64$, $N=128$, and $N=256$ both with and
without a bulk flow velocity of $v=10$ (Mach $M=6.9$) in the $y$-direction.  The mixing statistic
quantifies the qualitative impression that less mixing occurs in the simulations with
a large bulk flow, and that the vertical extent of the instabilities is less than in the
simulations without a bulk flow.  The instabilities grow at different rates depending
on the resolution, which occurs because perturbations with different wavenumbers $k$ grow
at different rates.  The characteristic KH instability growth time scales as $\tau \propto k^{-1}$,
so the instabilities in the highest resolution simulation (with larger available wavenumbers)
grow the fastest.  In the $N=128$ and $N=256$ simulations without bulk flows, the 
primary $n=4$ instabilities seeded in the initial conditions actually crest and meld
with smaller scale instabilities resulting from interactions with waves that
have traveled across the computational volume.  The cresting of the waves corresponds
to the decrease in the mixing statistic for the $N=128$ and $N=256$ near time $t=1$.
For the simulations with bulk flows, the
instability is greatly suppressed for $N=64$ but does develop at higher
resolutions. 
In the $N=128$ and $N=256$ simulations with bulk flows the development of the instability is somewhat slower 
than in the simulations without the bulk flows, but faster than in the static $N=64$
simulation. 

\begin{figure*}
\includegraphics[width=168mm]{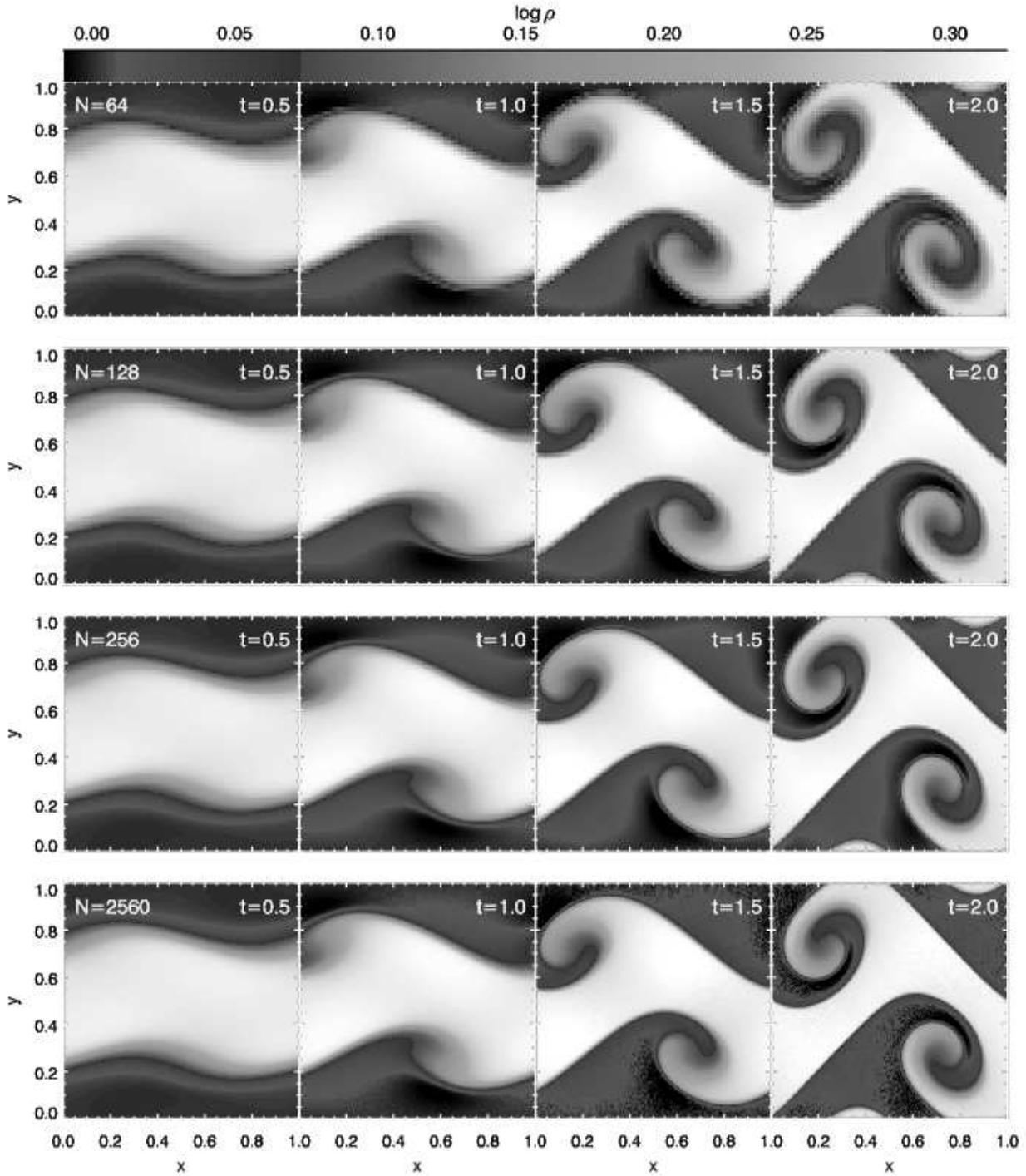}
\caption{
Kelvin-Helmholtz instability simulation of ICs B.  Shown is the temporal evolution of
the simulation with a mesh resolution of
$64\times64$ (first row), $128\times128$ (second row),  $256\times256$ (third row),
and $2560\times2560$ (fourth row) at times $t=0.5$ (first column), $t=1.0$ (second column), 
$t=1.5$ (third column), and $t=2.0$ (fourth column).
}
\label{fig:kh_abel_sequence}
\end{figure*} 
\begin{figure*}
\includegraphics[width=168mm]{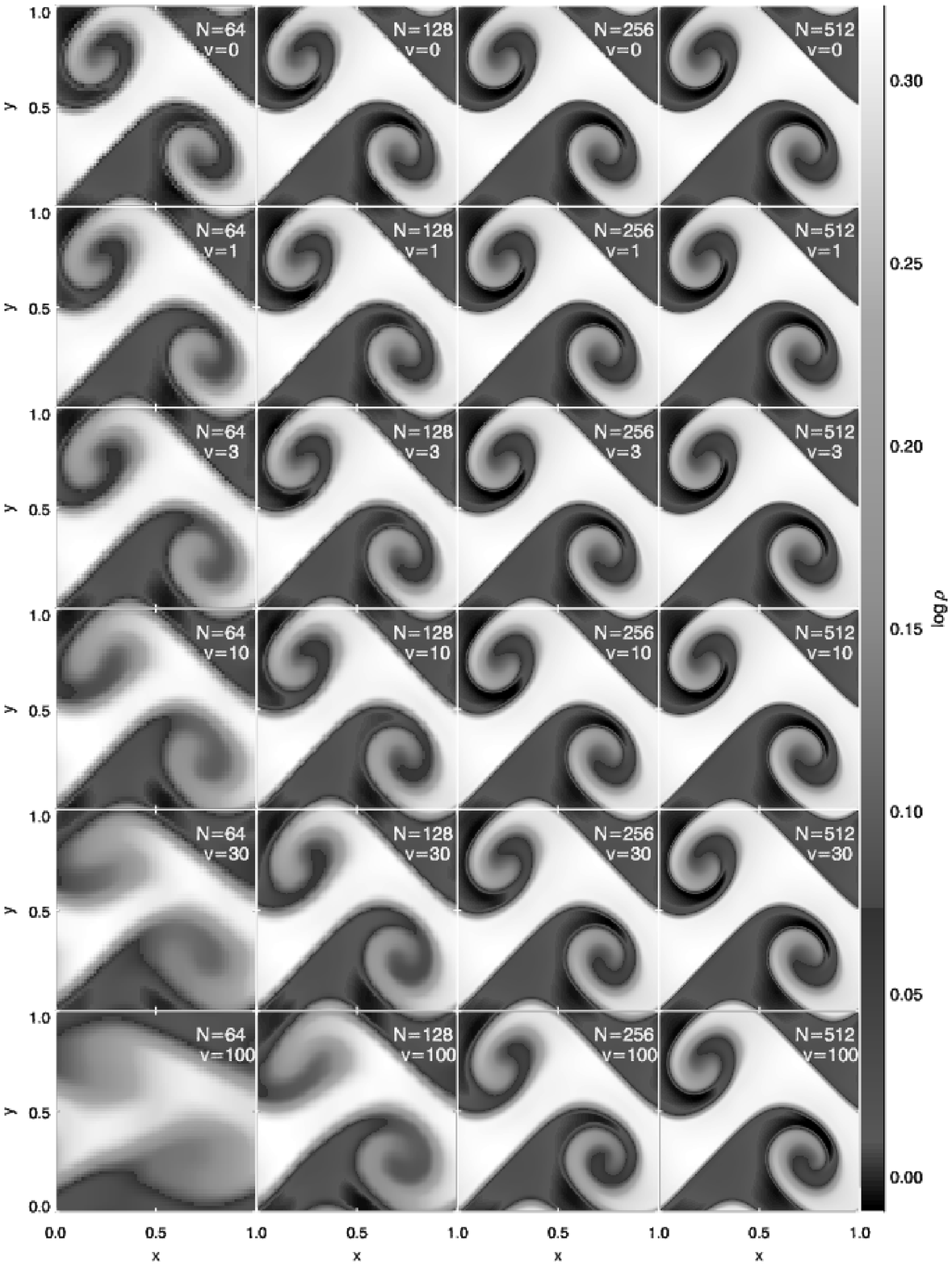}
\caption{
Kelvin-Helmholtz instability simulation of initial conditions ICs B at time $t=2$.
Shown is the simulation density distribution for grid resolutions of $N=64$ (first column),
$N=128$ (second column), $N=256$ (third column), and $N=512$ (fourth column).
Each grid resolution is simulated with bulk flow velocities of $v=[0,1,3,10,30,100]$ (Mach $M=[0,0.7,2.1,6.9,21,69]$, top-bottom rows).
The results of the $N=512$, $v=0$ run are used to define the $\Lone$ error norm for
this KH instability simulation.
}
\label{fig:final_sequence}
\end{figure*}

Given these results, one may ask is there a correct solution
to which the simulations should converge with increasing resolution for
any bulk velocity?  
The simulations shown in Figure
\ref{fig:kh_sequence} cover a factor of $40$ in resolution, and each
increase in resolution is followed by a corresponding increase in the
complexity of the small-scale structure of the instability.  Details
of the structures, however, are quite different in each case as is their
overall evolution shown in Fig.~\ref{fig:kh_abel_sequence}.  As
such, the solution does not converge with increasing resolution to any
well-defined configuration in the simulations with this setup. This result is
not surprising as the initial conditions with the sharp interface
allow all perturbation modes, both real and numerical, to grow
\citep{chandra1961a}. 
The modes excited by wave interactions or seeded by numerical noise depend on
the actual numerical resolution of the simulation and will be
different at different resolutions. This result is true not only for the static
mesh Eulerian calculations but also for the calculations with the moving mesh
code presented by \cite{springel2009a}. We therefore conclude that the
system ICs A cannot reliably 
be used for convergence studies or for the tests 
studying the development of a KH instability in the presence of a uniform bulk flow.

\subsection{Why Does The Simulation Evolution Depend On Bulk Velocity?}

The change in the evolution in the presence of
a bulk flow has been characterized as evidence for Galilean
non-invariance of the Eulerian methods \citep{springel2009a}.  However,
the cause of the differences has not been unambiguously identified.  First, as
stated above, the simulations without bulk flows do not converge with
increasing resolution.  The cause of this lack of convergence is the
excitation of small-scale modes by secondary waves driven by the
initially seeded $n=4$ perturbation.  These waves travel through the
low-density fluid, cross the computational volume, and interact with
the dense fluid.  The interaction between the waves and the
dense fluid drives high-frequency oscillations that quickly become
unstable.  At high resolution, numerical noise can contribute
additional small scale structure to these perturbations.  
These high-frequency modes can become unstable owing to the
sharp transition between the two fluids in the initial conditions.  If
these small-scale instabilities were suppressed, only the initially
seeded $n=4$ mode would grow.

In the simulations with a bulk flow, the sharp transition between the
two fluids is smeared owing to diffusive errors generated as fluid is 
advected through the grid --- an inherent property of all
Eulerian schemes, including those not based on Riemann solvers.
As discussed by \cite{chandra1961a}, the stability of Kelvin-Helmholtz
perturbations of different wavenumbers depends strongly on the density
and shearing velocity gradient present between the two fluids.  While
the connection between the instability of a given mode and the nature
of the gradient can be extremely complicated, as a rule of thumb in
the absence of gravity and surface tension shallower gradients lead to
an effective maximum unstable wavenumber of order the inverse of the
spatial scale of the gradient \citep[see the discussion in \S102 of
][]{chandra1961a}.  The numerical diffusion in the Eulerian scheme is
strong in the simulations with a large bulk flow and simply imposes
stability on small-scale perturbations.  In the simulations with large
bulk flows, only the initially seeded $n=4$ mode grows with time.  The
lowest resolution ($N=64$) simulation with a bulk flow has strong
enough numerical diffusion that the $n=4$ is not well resolved and
diffuses away before the shearing flow can cause the wave to crest.

The results of these Kelvin-Helmholtz simulations suggest that
numerical diffusion leads to change in the available modes that can
grow into instabilities for the chosen initial conditions.  The
approximation of the physical laws does not explicitly change, but the
error induced by numerical diffusion simply alters the physical system
being modeled.  While this new interpretation of the origin of the
differences in this Kelvin-Helmholtz simulation is straightforward, it
is unwieldly to test in this case because of the somewhat pathological
choice of initial conditions.  If the advection-related diffusion is
the origin of the ``Galilean non-invariance'' of the Eulerian schemes,
then the error of a numerical solution will depend on the bulk
velocity (because the integration to a given time will be carried out
with more time steps), but should decrease with increasing resolution.
We discuss this behavior further in \S~\ref{section:discussion}.  Since the error
norm of ICs A is ill-defined in the case without bulk flows a useful
error analysis would be difficult.  We will need to choose a different
set of initial conditions for a detailed error analysis of KH
instabilities in the presence of strong bulk flows.

\section{The Kelvin-Helmholtz Instability With A Gradual Interface}
\label{section:kh_abel}

As discussed in \S \ref{section:kh}, the numerical study of
Kelvin-Helmholtz instabilities can be complicated by the choice of
initial conditions.  If large wavenumber perturbations are unstable,
the growth of the instabilities can be strongly influenced by the
development of small scale modes seeded or affected by resolution.
As a result, the simulation may not converge with increasing resolution.
A reasonable solution to this problem is to choose the initial
conditions such that the stratification of the two fluids is not sharp
but gradual.  Such a setup approximates the interfaces that can arise in
simulations of real astrophysical systems where boundaries between fluids
are not perfectly discontinuous.  

We therefore alter the Kelvin-Helmholtz initial conditions 
from those used by \cite{springel2009a} through the use of a ``ramp'' function
\begin{equation}
\label{equation:ramp}
R(y) = \frac{1}{1 + \exp[ 2(y-0.25)/\Delta_{y}]}\frac{1}{1 + \exp[ 2(0.75-y)/\Delta_{y}]},
\end{equation}
\noindent
where we set the parameter $\Delta_{y}=0.05$.  The new density distribution is initialized to 
\begin{equation}
\label{equation:density_ramp}
\rho(y) = \rho_{1} + R(y)[\rho_{2} - \rho_{1}],
\end{equation}
\noindent
and the shearing velocity distribution is changed to 
\begin{equation}
\label{equation:velocity_ramp}
v_{x}(y) = v_{1} + R(y)[v_{2} - v_{1}].
\end{equation}
\noindent
The parameter values remain $\rho_{1}=2$,  $\rho_{2}=1$, $v_{1}=0.5$, 
and $v_{2}=0.5$, with a constant pressure $P=2.5$ and adiabatic index $\gamma=5/3$.
The initial velocity perturbation is set to 
\begin{equation}
\label{equation:abel_velocity_perturbation}
v_{y}(x) = w_{0} \sin(n\pi x),
\end{equation}
\noindent
with $w_{0} =0.1$ as before and $n=2$.  A lower frequency
perturbation is chosen to minimize the interaction between instabilities
after they become nonlinear, but we have checked that our conclusions
are not affected by this choice of perturbation (e.g., using $n=4$ 
leads to similar conclusions, see \S \ref{section:discussion} for
a discussion).  We will refer to these initial conditions as ``ICs B'',
and the corresponding density and shearing velocity distributions are compared with
ICs A in Figure \ref{fig:kh_ics}.

\begin{figure}
\includegraphics[width=84mm]{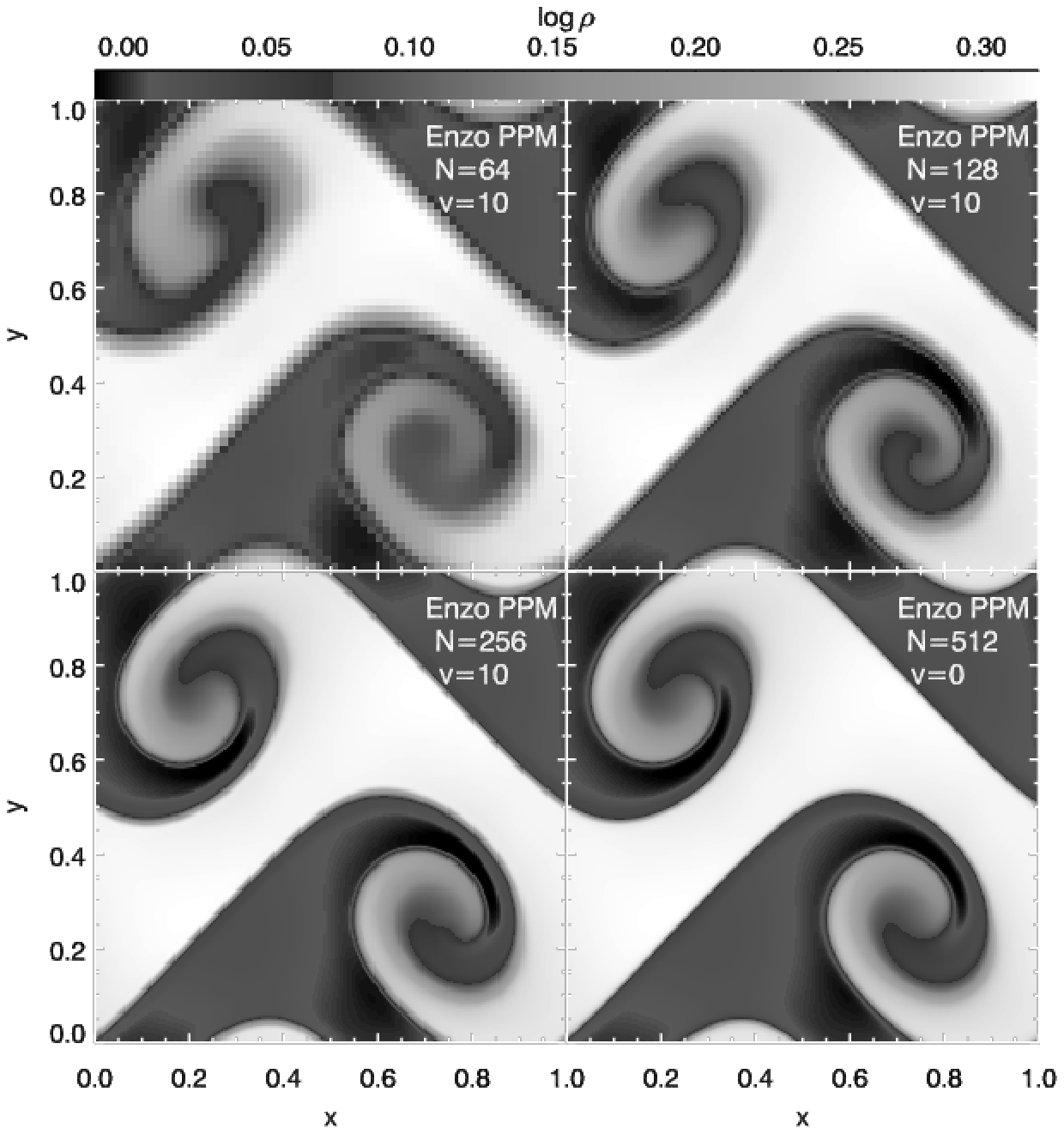}
\caption{
Kelvin-Helmholtz instability simulation performed with the Piecewise Parabolic Method version of the Enzo code (O'Shea et al. 2004).
Shown are the simulation results for the density at time $t=2$ for resolution and $y$-direction bulk velocities of $[N,v]=[64,10]$ (upper left panel),
$[N,v]=[128,10]$ (upper right panel), $[N,v]=[256,10]$ (lower left panel), and $[N,v]=[512,0]$ (lower right panel).  The error convergence
rate for PPM reconstruction, measured relative to the $[N,v]=[512,0]$ simulation, shows the expected improvement over the linear reconstruction results (see Figure \ref{fig:final_sequence}) as less diffusive methods should perform better in the presence of large advective flows.
}
\label{fig:enzo}
\end{figure} 
\begin{figure}
\includegraphics[width=84mm]{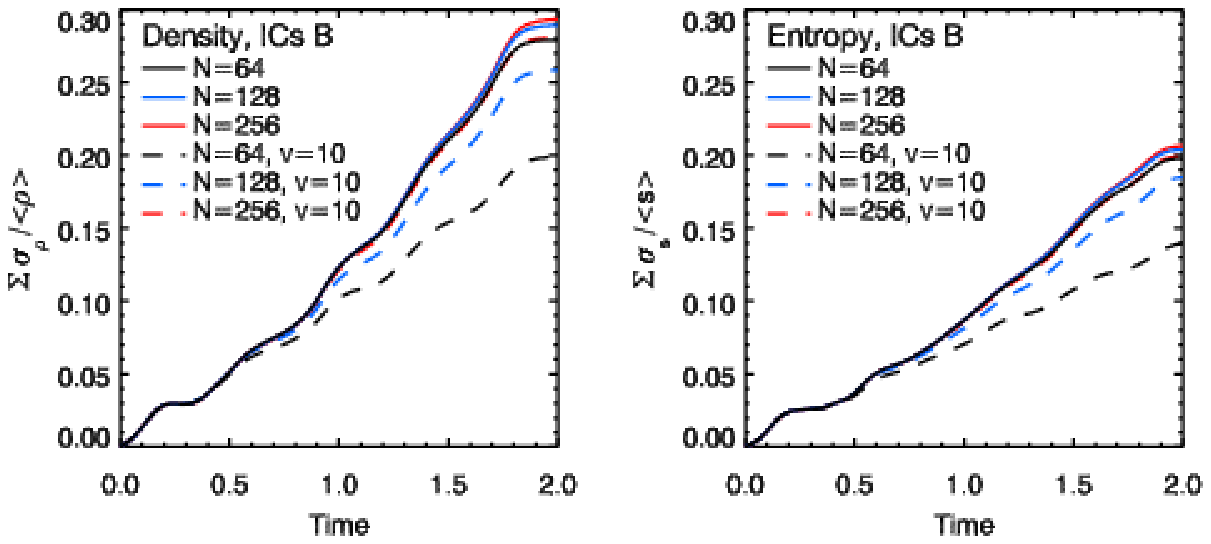}
\caption{
``Mixing statistic'' for the Kelvin-Helmholtz instability simulation of 
ICs B with time (see Equation \ref{equation:mixing_statistic}).  
Shown is a dimensionless measure of the root-mean-squared
density (left panel) and entropy (right panel) fluctuations from the growing
KH instabilities in simulations with $N=64$ (black), $N=128$ (blue), and
$N=256$ (red) resolution and without bulk flows, as well as $N=64$ (black dashed),
$N=128$ (blue dashed), and $N=256$ (red dashed) simulations with $v=10$ (Mach $M=6.9$) velocity bulk flow along the
$y$-direction.  In the simulations without bulk flows, the instabilities
grow at nearly the same rate. In the simulations with $v=10$
bulk flows, the instability growth improves with increasing resolution and
is comparable to the simulations with no bulk flow with a resolution $N=256$
or better.
}
\label{fig:kh_abel_statistics}
\end{figure}

\begin{figure}
\includegraphics[width=84mm]{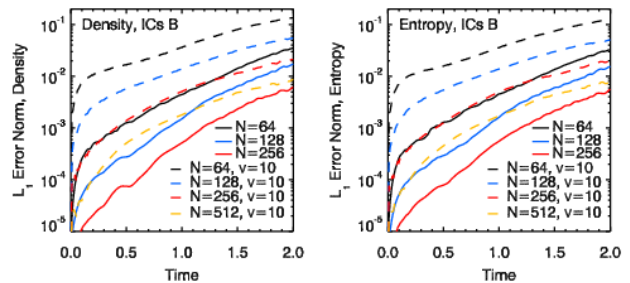}
\caption{
$\Lone$ error norm for the Kelvin-Helmholtz instability simulation of the
initial conditions ICs B with time.  Shown is the $\Lone$ error norm of simulations
with no bulk flow and grid resolutions $N=64$ (black), $N=128$ (blue), and $N=256$
(red), and simulations with bulk flow velocity $v=10$ (Mach $M=6.9$) with resolutions $N=64$ (black dashed),
$N=128$ (blue dashed), $N=256$ (red dashed), and $N=512$ (orange dashed).  In each case, the $\Lone$ error
norm is measured relative to a $N=512$ simulation with no bulk flow.  For a bulk flow
of velocity $v=10$, the effective resolution of the simulation is degraded by numerical
diffusion a factor $\sim4$ compared with simulations with no bulk flows.
}
\label{fig:kh_abel_error_norm}
\end{figure}

The inclusion of a finite gradient in the density and velocity distribution
in ICs B leads to a dramatic suppression of small scale features in the
growing KH instability.  Figure \ref{fig:kh_abel_sequence} shows the
temporal evolution of the KH instability arising from ICs B with no bulk flow, simulated
with grid resolutions of $N=64$ (first row), $N=128$ (second row), $N=256$ (third row), and
$N=2560$ (fourth row).  The density distribution of the computational 
volume is plotted at times $t=0.5$ (first column), $t=1.0$ (second column), 
$t=1.5$ (third column), and $t=2.0$ (fourth column), and is directly comparable
to the simulations of ICs A shown in Figure \ref{fig:kh_sequence}.
The evolution of the KH instability is
completely dominated by the seeded $n=2$ perturbation. 
As with ICs A, the velocity perturbation drives
secondary waves that cross the computational volume.  These waves
travel through the low density fluid and collide with the 
high density fluid after traversing the box.  In contrast to the
evolution of the instability in ICs A, these waves do not excite
other, higher frequency modes in the high density fluid.  The transition
region between the fluids oscillates after interacting with these waves,
but the density distribution is overstable at large wavenumbers and the
oscillations damp away.  As a result, the evolution of the KH instability
is nearly independent of the simulation resolution when no bulk
flow is included and converges to a well-defined solution.

In an attempt at a comprehensive study of the KH instability resulting
from ICs B, we perform a suite of 24 simulations with resolutions
$N=[64, 128, 256, 512]$, with each resolution simulation calculated
with bulk flow velocities of $v=[0,1,3,10,30,100]$ (Mach
$M=[0,0.7,2.1,6.9,21,69]$) in the $y$-direction.  The simulations were
performed in a manner identical to the simulations of ICs A, with the
Courant-Freidrichs-Lewy condition parameter $cfl=0.6$ and simulation
outputs recorded at time intervals of $\Delta t=0.01$.

Figure \ref{fig:final_sequence} shows the results of these 24 simulations
at time $t=2.0$, arrayed with resolution increasing to the right and
bulk flow velocity increasing from $v=0$ (top row) to $v=100$ (Mach $M=69$, bottom row).
The influence of a bulk flow on the evolution of the KH instability growing
from ICs B is much less dramatic than for ICs A.  The diffusive error induced
by the bulk flow has little influence on the physical development of the
instability, and only limits the growth of the instability for the lowest
resolution simulation ($N=64$) for bulk flow Mach numbers of $M\ga20$ 
($v\ga30$).  The diffusive error clearly decreases and the simulations
visually appear to converge with increasing resolution at each bulk
flow velocity.  The result is dramatic considering the extreme supersonic
bulk flow velocities (up to Mach number $M\sim70$) considered. Note that
for the run with the largest bulk velocity and the highest resolution, the
interface has been advected through $\approx 10^5$ computational cells. 

Since the degradation of the computed solution by numerical diffusion in the
presence of a bulk motion can be ameliorated by increasing the resolution, the
performance should also improve at fixed resolution when a higher-order method
is used.  The use of PPM reconstruction should then result in less diffusion than
when linear reconstruction is utilized.
To test
this intuition, we use the PPM version of the code Enzo 
\citep{bryan1997a,bryan1999a,norman1999a,bryan2001a,oshea2004a}, which 
uses an exact Riemann solver and \cite{strang1968a} dimensional splitting, and
perform exactly the same ramp KH instability simulation.  Figure \ref{fig:enzo}
shows the results for the density at time $t=2$ for simulations with resolutions
and $y$-direction bulk velocities of $N=[64,128,256,512]$ and $v=[10,10,10,0]$
using the Enzo PPM code with $cfl=0.8$.  First, the results of the Enzo PPM and ART simulations
are remarkably similar (Figures \ref{fig:final_sequence} and \ref{fig:enzo} are
directly comparable, with the same color scaling). 
Second, the final density distributions in the lower resolution 
($N=64,128,256$) simulations with $v=10$ bulk flows quickly converge with
increasing resolution to the reference high-resolution 
($[N,v]=[512,0]$) simulation results.  As expected, at fixed resolution the Enzo PPM results are
clearly less diffusive than the ART results (e.g., the $N=64,v=10$ simulation results in Figures \ref{fig:final_sequence}
and \ref{fig:enzo}).

\subsection{Error Analysis}
\label{section:error_analysis}

The apparent convergence of the simulation with increasing resolution for
each bulk flow can be quantified.  
Statistical measures of the instability evolution, including
both the $\Lone$ error norm (Equation \ref{equation:l1_error}) and the mixing statistic
(Equation \ref{equation:mixing_statistic}), are well-defined for the ICs B simulation
if one substitutes the results of a high-resolution simulation
for the ``true'' solution.  We adopt this approach and define the
error norm relative to a $512^2$ simulation with no bulk flow 
(upper right corner of Figure \ref{fig:final_sequence}), and
calculate all statistical measures by rebinning simulations to $64^2$
resolution when necessary.

For comparison with the results from
simulations of ICs A, Figure \ref{fig:kh_abel_statistics} shows the time evolution
of the density (left panel) and entropy (right panel) mixing statistics 
for the simulation of ICs B with resolutions
$N=[64, 128, 256]$ and bulk flow velocities $v=[0,10]$ (Mach $M=[0,6.9]$).
In dramatic contrast to results of ICs A, the mixing statistic for
ICs B is roughly independent of resolution for the simulations with
no bulk flow and clearly converges at resolution $N=256$ for a 
bulk flow velocity $v=10$.  Further, the instability grows in both 
simulations with and without a $v=10$ bulk flow for all the resolutions
studied.  For simulations with a $v=10$ bulk flow the growth rate of 
the instability changes with resolution, but the growth rate
agrees with the $v=0$ simulations by resolution $N=256$.

Figure \ref{fig:kh_abel_error_norm} shows the time evolution of the
$\Lone$ error norm for density (left panel) and entropy (right panel)
of the ICs B simulations with $N=[64,128,256]$ with no
bulk velocity ($v=0$, shown as black, blue, and red lines, respectively) and
simulations with $N=[64,128,256,512]$ and bulk velocity $v=10$ 
(shown as black, blue, red, and orange dashed lines, respectively).
As expected, the error declines rapidly with increasing resolution
and increases with increasing time of the simulation.  For the $v=10$ (Mach $M=6.9$)
bulk flow simulations the numerical diffusion degrades the effective
resolution of the simulation, with the $N=256$, $v=10$ simulation
performing comparably to the $N=64$, $v=0$ simulation and
the $N=512$, $v=10$ simulation performing comparably to the 
$N=128$, $v=0$ simulation.  

\begin{figure}
\includegraphics[width=84mm]{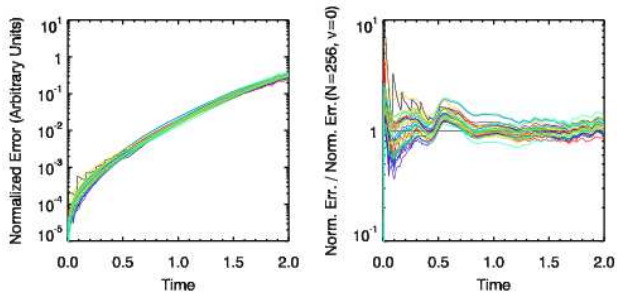}
\caption{
Common $\Lone$ error norm dependence on resolution and velocity.  Shown
is the $\Lone$ error norm for all 24 simulations in Figure \ref{fig:final_sequence}, normalized
by their common dependence on resolution ($\Lone\propto N^{-2}$) and velocity [$\Lone\propto(1+v)^{0.55}$]
(left panel).  This functional dependence is common to all the simulations (right panel), as demonstrated by
dividing out the time-dependence of the $N=256$, $v=0$ simulation described by Equation \ref{equation:l1_scaling}.
}
\label{fig:error_scaling}
\end{figure}

The time dependence of the $\Lone$
error norm in Figure \ref{fig:kh_abel_error_norm} appears to be self-similar 
for fixed bulk flow velocity.
Further, the logarithmic separation of
the $\Lone$ error norm for simulations with differing resolution appears
to be approximately constant.  With some experimentation, we
find that the $\Lone$ error norm of these simulations scales approximately
as
\begin{equation}
\label{equation:l1_scaling}
\Lone \propto N^{-2}(1+v)^{0.55}(1+t)^{[2(N/64)^{-0.5} + 2v^{0.06}]}.
\end{equation}
\noindent
Figure \ref{fig:error_scaling} shows the $\Lone$ error norm of all
24 simulations from Figure \ref{fig:final_sequence} normalized
by the scaling given by Equation \ref{equation:l1_scaling} (left panel;
the $512^2$, $v=0$ simulation error norm is $\Lone=0$ by definition).
The right panel of Figure \ref{fig:error_scaling} shows the normalized
$\Lone$ error norm curves from the left panel divided by the normalized
$\Lone$ error norm for the $N=256$, $v=0$ simulation.  
Figure \ref{fig:error_scaling} demonstrates that Equation 
\ref{equation:l1_scaling} accounts for the dependence of the $\Lone$
error norm on simulation resolution and velocity.  At fixed time,
the $\Lone$ error norm dependence on resolution scales as 
$\Lone\propto N^{-2}$ as expected for the spatially 
second-order accurate Eulerian method used by ART.  The velocity
dependence of the error norm is $\Lone\propto(1+v)^{0.55}$, which
is remarkably similar to the $\Lone\propto v^{0.5}$ expected from
numerical diffusion.  We therefore suggest that the 
velocity-dependence of the $\Lone$ error norm at fixed time 
is consistent with numerical diffusion alone.  
The time-dependence
of the $\Lone$ error norm scaling has only a very weak apparent
dependence on velocity $\Lone\propto(1+t)^{2v^{0.06}}$, and a
moderate dependence on the resolution as $\Lone\propto(1+t)^{2(N/64)^{0.5}}$.
For a given bulk flow velocity, Equation \ref{equation:l1_scaling} can be used to
calculate the necessary resolution required in simulations of
this KH instability\footnote{The scaling could conceivably be different in other 
simulation problems.} to reach an equivalent error for the same simulation
without a bulk flow. For the setup of ICs B, although results 
of runs with and without bulk velocity have different errors (expectedly, given
the different number of time steps and different amount of advection), calculations
converge to the same Galilean invariant result as the resolution is 
increased.

The same error analysis can be performed for the Enzo results as for the
ART results above.  We use the $N=512$, $v=0$ Enzo PPM simulation, rebinned to $N=64$ resolution, to define the
reference solution in Equation \ref{equation:l1_error}.  The
$\Lone$ error norms are then measured for the $N=[64,128,256]$, $v=10$ Enzo PPM simulations.
We find that at early times the error norm improves with resolution faster than quadratically,
as expected for a PPM code.
Compared with the results from the ART code,
these error convergence rates are consistent with the precision increase afforded by the PPM reconstruction.
We therefore suggest that the comparison of the ART and Enzo PPM results demonstrate that both codes
converge correctly with increasing resolution for their formal order. 
We therefore expect that other Eulerian
codes will also produce Galilean invariant solutions to within a specified error tolerance for similar hydrodynamical
problems.

\section{Discussion}
\label{section:discussion}

The simulations of KH instabilities presented in this paper suggest
that the apparent Galilean non-invariance in Eulerian hydrodynamical
calculations owes to numerical diffusion associated with advection
of the fluid through the computational grid.
The diffusive
errors arise from the truncation errors associated with
discretization of spatial and temporal derivatives in the computational method.  The
traditional meaning of Galilean invariance is that the formulation of
a specified physical law in two different inertial frames is related
by a Galilean transformation.  While the Euler equations are
Galilean-invariant, discretized approximations to the Euler equations
are not guaranteed to obey the same transformational properties.
However, with increasing spatial resolution simulations converge to
the Galilean invariant numerical solution (at least for 
problems that have convergent solutions).

In KH instability simulation initial conditions with discontinuously 
sharp boundaries between shearing fluids,
all modes down to the cell scale are unstable.
Secondary waves driven by the seeded perturbation and numerical
noise can then generate small-scale instabilities that affect 
the long-term evolution of the system. 
For such initial conditions, numerical diffusion can both alter the 
frequency range of unstable wavenumbers and smear out small-scale features
in the flow.  In the presence of a large bulk flow, the increased
numerical diffusion can thereby suppress the growth of small-scale instabilities
through these effects. In such tests Eulerian
methods can produce qualitatively different results compared with simulations
of the same initial conditions that do not include a bulk flow.
We emphasize that these differences arise 
because the development of both real and 
numerical small-scale perturbations
in the flow are affected by diffusive errors associated with advection
through the grid. 

 For physical systems in which all of the perturbations are well-resolved,
 our results show that calculations quickly converge to a well-defined, Galilean invariant
solution. This result holds even in the presence of
highly supersonic bulk motions, although in this case spatial resolution
needs to scale approximately as $N\propto v^{0.5}$ to counteract increased diffusive
errors owing to advection.
Supersonic bulk flows therefore pose no serious problem
for Eulerian simulations as convergence studies -- the final arbiter of
any quantitatively credible numerical calculation -- will allow one to
test for a converged solution regardless of the magnitude of the bulk flow
for realistic initial conditions.  
Although the required increase of resolution may seem like a large price 
to pay,\footnote{Note that the use of moving mesh to handle the bulk flows is also
expensive, as it requires both large memory per cell and a fairly sophisticated
algorithm for handling the motion of the mesh and regridding.}
 adaptive mesh refinement makes it considerably easier to achieve 
such resolution increases in the relevant regions around the fluid interface. 
We have calculated some of the tests presented in the paper with adaptive
mesh refinement using refinement conditions based
on density and entropy gradients and obtained results similar to those
achieved with a uniform grid of higher resolution.  

We caution that numerical diffusion may limit what physical systems can be
accurately modeled by Eulerian methods with arbitrary flow
velocities if the resolution is poor.  Since numerical diffusion can alter the physical system
being modeled (for instance, in the presence of unrealistic discontinuities), 
simulation results 
in the Eulerian method can depend irrevocably on the local flow velocity in fixed, 
low-resolution calculations.
Our studies suggest that 
hydrodynamical systems altered by numerical diffusion (like ICs A) are
Galilean invariant to within a specified error tolerance, but these
calculations would model a different system in the effective absence 
of numerical diffusion (since, e.g., the frequency range of unstable modes
has changed).  
The failure of KH instabilities to develop in low-resolution simulations
with large bulk velocities can be directly rectified with sufficient resolution.
We therefore suggest that knowledge of the bulk velocity, local density and shear
velocity gradients, or other gauges of the potential impact of numerical diffusion
be incorporated into
resolution criteria and, for adaptive mesh codes, into refinement conditions. 

When Eulerian methods are used to model the growth of instabilities in the
presence of a bulk velocity, care should be exercised to ensure that
the perturbations of interest are not made overstable by numerical diffusion
at low resolution.
For instance, if the perturbation frequency $n$ in Equation 
\ref{equation:abel_velocity_perturbation} was greatly increased such that the 
perturbation wavenumber was close to the maximum unstable wavenumber permitted
by the system, then
numerical diffusion could, in principle, artificially alter the growth of
the perturbation if the mode was not properly resolved.  We have repeated simulations of ICs B with $n=4$ and
find similar convergence properties to those reported for $n=2$.  However,
the typical error norm at fixed resolution and velocity is larger for $n=4$ than for
$n=2$.  Sensibly, smearing in the solution owing to numerical diffusion will
affect small scales more so than large scale modes.

\subsection{Limiting Numerical Diffusion}
\label{section:numerical_diffusion}

Techniques exist for reducing numerical diffusion in Eulerian methods, and
an appropriate application of such techniques could help mitigate 
velocity-dependent behavior of calculated solutions.  We demonstrated that
intrinsically less
diffusive Eulerian methods, such as PPM, show improved results
over codes using linear reconstruction.  The ART simulations presented in this
paper use the \cite{van_leer1977a} slope limiter, but the use of less-diffusive slope
limiters during reconstruction, such as the Superbee limiter \citep{roe1981a},
improve results for some physical systems.  
Artificial compression in the form of slope steepeners might also help for
some problems.  
For instance, our tests presented in \S \ref{section:diffusion} show that
the one dimensional advection of square
wave contact discontinuities can be made insensitive to even ultrasonic (Mach $M\sim10^{4}$) advective
velocities if the 
the slope steepening method of \citet[][]{yang1990a} is used.
However, when applied to multidimensional
problems, such as the KH instability simulations presented here, we have found that the
\cite{yang1990a} slope steepener produces oddly angular features in the 
fluid flow when used with a \cite{strang1968a} dimensionally split solver (we
have not tested slope steepeners with unsplit solvers).  Similarly, the \cite{harten1989a}
subcell resolution method produces excellent results in one dimension but is
unwieldly in multiple dimensions.  Other approaches for reducing numerical
diffusion in Eulerian methods are discussed at length by \cite{laney1998a}.

The strength of numerical diffusion can differ on the upwind and downwind
side of an initially symmetrical waveform being advected using Godunov-type
Eulerian codes.  The results of \S \ref{fig:kh_vel_sequence}
show clear evidence for
larger numerical diffusion in the high-to-low density transition 
than for the low-to-high density transition, and for an alteration of
the piecewise reconstruction of the initially symmetrical waveform by numerical diffusion.
It should be noted that some reconstruction methods intrinsically do not 
produce symmetrical
piecewise approximations to smooth functions, depending on the number of
samples of the waveform.  For instance, average-quadratic interpolations 
of sine and square waves 
using
essentially nonoscillatory reconstruction 
are
not always symmetrical \citep[see \S 9.2 of][]{laney1998a}.  Numerical
diffusion may depend on the flow velocity and direction in such cases.

\subsection{Gravity and the Kelvin-Helmholtz Instability}
\label{section:kh_gravity}

While our study has focused on idealized hydrodynamical systems, 
real astrophysical systems where the Kelvin-Helmholtz instability 
operates will be influenced by gravitational fields.  Mixing in
a stratified fluid in the presence of gravity must overcome
gravitational potential energy with kinetic energy from the shearing
motion.  The stabililty of the fluid against Kelvin-Helmholtz modes
can therefore be characterized by the Richardson number that measures
the relative strength of buoyancy in the fluid and inertia supplied by
the shearing velocity gradient $\mathrm{d}v/\mathrm{d}z$ as
\begin{equation}
\label{equation:richardson_numbeR}
J = -\frac{g}{\rho}\frac{\mathrm{d}\rho/\mathrm{d}z}{(\mathrm{d}v/\mathrm{d}z)^{2}},
\end{equation}
\noindent
where $g$ is the local gravitational acceleration along the $z$-direction and
$\rho$ and $\mathrm{d}\rho/\mathrm{d}z$ are the local density and density gradient 
in the fluid.  \cite{chandra1961a} shows that since the kinetic energy of the shearing
motion powers the instability, if the shearing kinetic energy is too small to overcome the
gravitational potential energy of the fluid then the KH instability will not occur.
In terms of the Richardson number, it is straightforward to show that the corresponding
necessary (but not sufficient) condition for stability is $J>1/4$.\footnote{In the absence of gravity $J=0$, and 
stability depends on whether individual modes can be excited.  In the presence of
other effects, such as a Coriolis force, instability may require higher Richardson
numbers \citep[see, e.g.,][]{gomez2005a}.}

In the context of modeling astrophysical systems, our results suggest that 
numerical diffusion may act to change the Richardson number by 
altering the local density and shear velocity gradients.  Since the
numerical diffusion will smooth density and velocity gradients by similar amounts,
the effect of numerical diffusion will typically be to increase the stability of
KH modes.  Simulations of systems with modes that have $J\sim1/4$ should therefore 
be checked to gauge the influence of any artificial stabilization from numerical
diffusion on the global
evolution of the system.

\section{Summary}
\label{section:summary}

In this paper we have presented hydrodynamical simulations of
Kelvin-Helmholtz (KH) instabilities to study the behavior of the
numerical solution in the presence of uniform bulk motion. We perform
these calculations to verify and evaluate previous claims in the
literature that Godunov-type Eulerian mesh calculations are
inherently Galilean non-invariant.

The KH instability study of \cite{springel2009a} was performed using
the ART \citep{kravtsov1997a,kravtsov_etal02} hydrodynamical code over a range of
numerical resolutions and bulk flow velocities.  We confirm the
results of \cite{springel2009a} that for low-resolution simulations
KH instabilities may not develop in the presence of bulk flows, but
show that such instabilities do develop with sufficient resolution.
We also explain why the \cite{springel2009a} KH instability simulation 
results generally depend on whether a bulk flow is
included.  
Diffusion in the presence of a bulk flow softens the sharp discontinuity 
between fluids in these initial conditions, thereby changing the frequency
range of unstable modes.  As a result, the small-scale structure of the
solution changes significantly with resolution and in the presence of large
advective velocity.
However, we emphasize that this velocity dependence owes to numerical
diffusion associated with advection of the fluid,
not because the analytical Riemann solution is somehow Galilean non-invariant (it is invariant).

We test this explanation by simulating another Kelvin-Helmholtz instability that
grows from initial conditions that have a gradual interface between fluids, in contrast to
the sharp interface used in the \cite{springel2009a} KH test.  In the new KH simulation, 
the stability of the fluid to growing small-scale modes allows
the seeded $n=2$ perturbation to develop unimpeded as a KH instability, and the
simulation demonstrates convergence with increasing resolution.  Furthermore,
the development of the KH instability occurs in the same manner for all bulk
flows examined (including supersonic bulk flows with Mach numbers $M\sim70$).
An analysis of the $\Lone$ error norm suggests that numerical diffusion accounts for
the entire error budget of the simulations that include bulk flows.  
We support this conclusion by demonstrating that the intrinsically less diffusive
Piecewise Parabolic Method  code Enzo 
\citep{bryan1997a,bryan1999a,norman1999a,bryan2001a,oshea2004a}
exhibits more rapid convergence 
than codes using linear reconstruction, and that the simulation results produced
by ART and Enzo are consistent.
For this
KH instability, the Galilean non-invariance can therefore be entirely accounted for by
numerical diffusion and can be effectively eliminated 
with increasing numerical resolution.

Our results suggest that physical systems where numerical
diffusion does not significantly alter the frequency range of unstable modes,
Godunov-type Eulerian methods will be Galilean invariant to within a
specified numerical error and that this error will decrease with 
increasing resolution.
We have demonstrated this result explicitly in the
case of a KH instability, but we suspect our conclusions will generalize to
other hydrodynamical instabilities.  Similar conclusions can be drawn from
test calculations by other authors \citep[e.g., the Gresho vortex tests by][]{springel2009a}.
These results show that there is no generic problem of using the Eulerian methods
for modeling complicated astrophysical systems with large bulk flows.
However, overcoming diffusive errors will require more stringent resolution
when modeling systems with large bulk fluid motions.

As with most numerical calculations, convergence studies are essential
for evaluating how numerical diffusion introduces velocity-dependent
error into the presented simulations.  If a full convergence study is
difficult or impossible, such as for the \cite{springel2009a} KH
instability simulation, caution needs to be exercised in interpreting
results in the presence of high-Mach number flows.

\section*{Acknowledgements}
BER would like to thank the staff of the
Neonatal Intensive Care Unit at the 
Comer Childrens' Hospital and the Mitchell
Transitional Care Unit at the University of
Chicago Medical Center for their hospitality
while this work was completed.
We also thank Anatoly Klypin, Brian O'Shea,
Eve Ostriker, 
Volker Springel, and Romain Teyssier
for helpful discussions.
BER gratefully acknowledges support from a
Spitzer Fellowship through a NASA grant
administrated by the Spitzer Science Center
during the majority of this work.
AVK is supported by the NSF under grants 
AST-0239759 and AST-0507666 and by NASA
through grant NAG5-13274.  BER and AVK
were also partially supported by the Kavli
Institute for Cosmological Physics at the 
University of Chicago.  
D.H.R. gratefully acknowledges the support of the
Institute for Advanced Study and the NSF through
grant AST-0807444.
Some of the simulations used in this work 
have been performed on the Joint 
Fermilab - KICP Supercomputing Cluster, 
supported by grants from Fermilab, Kavli 
Institute for Cosmological Physics, and 
the University of Chicago.


\begin{thebibliography}{}

\bibitem[\protect\citeauthoryear{{Agertz}, {Moore}, {Stadel}, {Potter},
  {Miniati}, {Read}, {Mayer}, {Gawryszczak}, {Kravtsov}, {Nordlund}, {Pearce},
  {Quilis}, {Rudd}, {Springel}, {Stone}, {Tasker}, {Teyssier}, {Wadsley} \&
  {Walder}}{{Agertz} et~al.}{2007}]{agertz2007a}
{Agertz} O.,  {Moore} B.,  {Stadel} J.,  {Potter} D.,  {Miniati} F.,  {Read}
  J.,  {Mayer} L.,  {Gawryszczak} A.,  {Kravtsov} A.,  {Nordlund} {\AA}.,
  {Pearce} F.,  {Quilis} V.,  {Rudd} D.,  {Springel} V.,  {Stone} J.,  {Tasker}
  E.,  {Teyssier} R.,  {Wadsley} J.,    {Walder} R.,  2007, MNRAS, 380, 963

\bibitem[\protect\citeauthoryear{{Balsara}}{{Balsara}}{1998}]{balsara1998a}
{Balsara} D.~S.,  1998, ApJS, 116, 133

\bibitem[\protect\citeauthoryear{{Bland-Hawthorn}, {Sutherland}, {Agertz} \&
  {Moore}}{{Bland-Hawthorn} et~al.}{2007}]{bland-hawthorn2007a}
{Bland-Hawthorn} J.,  {Sutherland} R.,  {Agertz} O.,    {Moore} B.,  2007,
  ApJL, 670, L109

\bibitem[\protect\citeauthoryear{{Boris} \& {Book}}{{Boris} \&
  {Book}}{1973}]{boris1973a}
{Boris} J.~P.,  {Book} D.~L.,  1973, Journal of Computational Physics, 11, 38

\bibitem[\protect\citeauthoryear{{Bryan}}{{Bryan}}{1999}]{bryan1999a}
{Bryan} G.~L.,  1999, Comput.~Sci.~Eng., Vol.~1, No.~2, p.~46 - 53, 1, 46

\bibitem[\protect\citeauthoryear{{Bryan}, {Abel} \& {Norman}}{{Bryan}
  et~al.}{2001}]{bryan2001a}
{Bryan} G.~L.,  {Abel} T.,    {Norman} M.~L.,  2001, ArXiv Astrophysics
  e-prints

\bibitem[\protect\citeauthoryear{{Bryan}, {Cen}, {Norman}, {Ostriker} \&
  {Stone}}{{Bryan} et~al.}{1994}]{bryan_etal94}
{Bryan} G.~L.,  {Cen} R.,  {Norman} M.~L.,  {Ostriker} J.~P.,    {Stone} J.~M.,
   1994, ApJ, 428, 405

\bibitem[\protect\citeauthoryear{{Bryan} \& {Norman}}{{Bryan} \&
  {Norman}}{1997}]{bryan1997a}
{Bryan} G.~L.,  {Norman} M.~L.,  1997, ArXiv Astrophysics e-prints

\bibitem[\protect\citeauthoryear{{Bryan}, {Norman}, {Stone}, {Cen} \&
  {Ostriker}}{{Bryan} et~al.}{1995}]{bryan_etal95}
{Bryan} G.~L.,  {Norman} M.~L.,  {Stone} J.~M.,  {Cen} R.,    {Ostriker} J.~P.,
   1995, Computer Physics Communications, 89, 149

\bibitem[\protect\citeauthoryear{{Cen}, {Ostriker}, {Jameson} \& {Liu}}{{Cen}
  et~al.}{1990}]{cen_etal90}
{Cen} R.~Y.,  {Ostriker} J.~P.,  {Jameson} A.,    {Liu} F.,  1990, ApJL, 362,
  L41

\bibitem[\protect\citeauthoryear{{Chandrasekhar}}{{Chandrasekhar}}{1961}]{chan%
dra1961a}
{Chandrasekhar} S.,  1961, {Hydrodynamic and hydromagnetic stability.}.
Dover Publications, NY, NY

\bibitem[\protect\citeauthoryear{{Clarke}}{{Clarke}}{1996}]{clark1996a}
{Clarke} D.~A.,  1996, ApJ, 457, 291

\bibitem[\protect\citeauthoryear{{Colella} \& {Glaz}}{{Colella} \&
  {Glaz}}{1985}]{colella1985a}
{Colella} P.,  {Glaz} H.~M.,  1985, Journal of Computational Physics, 59, 264

\bibitem[\protect\citeauthoryear{{Colella} \& {Woodward}}{{Colella} \&
  {Woodward}}{1984}]{colella1984a}
{Colella} P.,  {Woodward} P.~R.,  1984, Journal of Computational Physics, 54,
  174

\bibitem[\protect\citeauthoryear{{Courant}, {Isaacson} \& {Rees}}{{Courant}
  et~al.}{1952}]{courant1952a}
{Courant} R.,  {Isaacson} E.,    {Rees} M.,  1952, Comm. Pure Appl. Math, 5,
  243

\bibitem[\protect\citeauthoryear{{Fryxell}, {M{\"u}ller} \& {Arnett}}{{Fryxell}
  et~al.}{1989}]{fryxell_etal89}
{Fryxell} B.,  {M{\"u}ller} E.,    {Arnett} D.,  1989, in {Hillebrandt} W.,
  {M{\"u}ller} E.,  eds, Nuclear Astrophysics {Computation of multi-dimensional
  flows with non-uniform composition.}.
pp 100--102

\bibitem[\protect\citeauthoryear{{Fryxell}, {Olson}, {Ricker}, {Timmes},
  {Zingale}, {Lamb}, {MacNeice}, {Rosner}, {Truran} \& {Tufo}}{{Fryxell}
  et~al.}{2000}]{fryxell2000a}
{Fryxell} B.,  {Olson} K.,  {Ricker} P.,  {Timmes} F.~X.,  {Zingale} M.,
  {Lamb} D.~Q.,  {MacNeice} P.,  {Rosner} R.,  {Truran} J.~W.,    {Tufo} H.,
  2000, ApJS, 131, 273

\bibitem[\protect\citeauthoryear{{Gardiner} \& {Stone}}{{Gardiner} \&
  {Stone}}{2008}]{gardiner2008a}
{Gardiner} T.~A.,  {Stone} J.~M.,  2008, Journal of Computational Physics, 227,
  4123

\bibitem[\protect\citeauthoryear{{Godunov}}{{Godunov}}{1959}]{godunov1959a}
{Godunov} S.,  1959, Math. Sbornik, 47, 271

\bibitem[\protect\citeauthoryear{{G{\'o}mez} \& {Ostriker}}{{G{\'o}mez} \&
  {Ostriker}}{2005}]{gomez2005a}
{G{\'o}mez} G.~C.,  {Ostriker} E.~C.,  2005, ApJ, 630, 1093

\bibitem[\protect\citeauthoryear{{Harten}}{{Harten}}{1983}]{harten1983a}
{Harten} A.,  1983, Journal of Computational Physics, 49, 357

\bibitem[\protect\citeauthoryear{{Harten}}{{Harten}}{1989}]{harten1989a}
{Harten} A.,  1989, Journal of Computational Physics, 83, 148

\bibitem[\protect\citeauthoryear{{Hayes}, {Norman}, {Fiedler}, {Bordner}, {Li},
  {Clark}, {ud-Doula} \& {Mac Low}}{{Hayes} et~al.}{2006}]{hayes2006a}
{Hayes} J.~C.,  {Norman} M.~L.,  {Fiedler} R.~A.,  {Bordner} J.~O.,  {Li}
  P.~S.,  {Clark} S.~E.,  {ud-Doula} A.,    {Mac Low} M.-M.,  2006, ApJS, 165,
  188

\bibitem[\protect\citeauthoryear{{Heitsch} \& {Putman}}{{Heitsch} \&
  {Putman}}{2009}]{heitsch2009a}
{Heitsch} F.,  {Putman} M.~E.,  2009, ApJ, 698, 1485

\bibitem[\protect\citeauthoryear{{Helmholtz}}{{Helmholtz}}{1868}]{helmholtz186%
8a}
{Helmholtz} H.,  1868, Monthly Reports of the Royal Prussian Academy of
  Philosophy in Berlin, 23, 215

\bibitem[\protect\citeauthoryear{{Kelvin}}{{Kelvin}}{1910}]{kelvin1910a}
{Kelvin} W.,  1910, Mathematical and Physical Papers, 4

\bibitem[\protect\citeauthoryear{{Khokhlov}}{{Khokhlov}}{1998}]{khokhlov98}
{Khokhlov} A.,  1998, Journal of Computational Physics, 143, 519

\bibitem[\protect\citeauthoryear{{Kravtsov}, {Klypin} \& {Hoffman}}{{Kravtsov}
  et~al.}{2002}]{kravtsov_etal02}
{Kravtsov} A.~V.,  {Klypin} A.,    {Hoffman} Y.,  2002, ApJ, 571, 563

\bibitem[\protect\citeauthoryear{{Kravtsov}, {Klypin} \& {Khokhlov}}{{Kravtsov}
  et~al.}{1997}]{kravtsov1997a}
{Kravtsov} A.~V.,  {Klypin} A.~A.,    {Khokhlov} A.~M.,  1997, ApJS, 111, 73

\bibitem[\protect\citeauthoryear{{Laney}}{{Laney}}{1998}]{laney1998a}
{Laney} C.~B.,  1998, {Computational Gasdynamics.}.
Cambridge University Press

\bibitem[\protect\citeauthoryear{{Liu}, {Osher} \& {Chan}}{{Liu}
  et~al.}{1994}]{liu1994a}
{Liu} X.-D.,  {Osher} S.,    {Chan} T.,  1994, Journal of Computational
  Physics, 115, 200

\bibitem[\protect\citeauthoryear{{Mori} \& {Burkert}}{{Mori} \&
  {Burkert}}{2000}]{mori2000a}
{Mori} M.,  {Burkert} A.,  2000, ApJ, 538, 559

\bibitem[\protect\citeauthoryear{{Murray}, {White}, {Blondin} \&
  {Lin}}{{Murray} et~al.}{1993}]{murray1993a}
{Murray} S.~D.,  {White} S.~D.~M.,  {Blondin} J.~M.,    {Lin} D.~N.~C.,  1993,
  ApJ, 407, 588

\bibitem[\protect\citeauthoryear{{Norman} \& {Bryan}}{{Norman} \&
  {Bryan}}{1999}]{norman1999a}
{Norman} M.~L.,  {Bryan} G.~L.,  1999, in {S.~M.~Miyama, K.~Tomisaka, \&
  T.~Hanawa} ed., Numerical Astrophysics Vol.~240 of Astrophysics and Space
  Science Library, {Cosmological Adaptive Mesh Refinement}.
pp 19--+

\bibitem[\protect\citeauthoryear{{O'Shea}, {Bryan}, {Bordner}, {Norman},
  {Abel}, {Harkness} \& {Kritsuk}}{{O'Shea} et~al.}{2004}]{oshea2004a}
{O'Shea} B.~W.,  {Bryan} G.,  {Bordner} J.,  {Norman} M.~L.,  {Abel} T.,
  {Harkness} R.,    {Kritsuk} A.,  2004, ArXiv Astrophysics e-prints

\bibitem[\protect\citeauthoryear{{Park} \& {Schowengerdt}}{{Park} \&
  {Schowengerdt}}{1983}]{park1983a}
{Park} S.~K.,  {Schowengerdt} R.~A.,  1983, Computer Graphics Image Processing,
  23, 258

\bibitem[\protect\citeauthoryear{{Plewa} \& {M{\"u}ller}}{{Plewa} \&
  {M{\"u}ller}}{2001}]{plewa_mueller01}
{Plewa} T.,  {M{\"u}ller} E.,  2001, Computer Physics Communications, 138, 101

\bibitem[\protect\citeauthoryear{{Quilis}}{{Quilis}}{2004}]{quilis04}
{Quilis} V.,  2004, MNRAS, 352, 1426

\bibitem[\protect\citeauthoryear{{Quilis}, {Ibanez} \& {Saez}}{{Quilis}
  et~al.}{1996}]{quilis_etal96}
{Quilis} V.,  {Ibanez} J.~M.~A.,    {Saez} D.,  1996, ApJ, 469, 11

\bibitem[\protect\citeauthoryear{{Quilis}, {Moore} \& {Bower}}{{Quilis}
  et~al.}{2000}]{quilis_etal00}
{Quilis} V.,  {Moore} B.,    {Bower} R.,  2000, Science, 288, 1617

\bibitem[\protect\citeauthoryear{{Quirk}}{{Quirk}}{1994}]{quirk94}
{Quirk} J.~J.,  1994, Int.J. for Num. Methods in Fluids, 18, 555

\bibitem[\protect\citeauthoryear{{Quirk}}{{Quirk}}{2005}]{quirk05}
{Quirk} J.~J.,  2005, Adaptive Mesh Refinement, theory and applications,
  Springer-Verlag, T. Plewa, T.J.Linde, V. Gregory, eds., pp 3--28

\bibitem[\protect\citeauthoryear{{Ricker}, {Dodelson} \& {Lamb}}{{Ricker}
  et~al.}{2000}]{ricker_etal00}
{Ricker} P.~M.,  {Dodelson} S.,    {Lamb} D.~Q.,  2000, ApJ, 536, 122

\bibitem[\protect\citeauthoryear{{Roe}}{{Roe}}{1981}]{roe1981a}
{Roe} P.~L.,  1981, Journal of Computational Physics, 43, 357

\bibitem[\protect\citeauthoryear{{Roe} \& {Pike}}{{Roe} \&
  {Pike}}{1984}]{roe1984a}
{Roe} P.~L.,  {Pike} J.,  1984, in {Glowinski} R.,  {Lions} J.-L.,  eds,
  Computing Methods in Applied Sciences and Engineering Vol.~6, {Efficient
  Construction and Utilization of Approximate Riemann Solutions}.
pp 499--518

\bibitem[\protect\citeauthoryear{{Ryu}, {Ostriker}, {Kang} \& {Cen}}{{Ryu}
  et~al.}{1993}]{ryu1993a}
{Ryu} D.,  {Ostriker} J.~P.,  {Kang} H.,    {Cen} R.,  1993, ApJ, 414, 1

\bibitem[\protect\citeauthoryear{{Springel}}{{Springel}}{2009}]{springel2009a}
{Springel} V.,  2009, ArXiv e-prints

\bibitem[\protect\citeauthoryear{{Stone}, {Gardiner}, {Teuben}, {Hawley} \&
  {Simon}}{{Stone} et~al.}{2008}]{stone2008a}
{Stone} J.~M.,  {Gardiner} T.~A.,  {Teuben} P.,  {Hawley} J.~F.,    {Simon}
  J.~B.,  2008, ApJS, 178, 137

\bibitem[\protect\citeauthoryear{{Stone} \& {Norman}}{{Stone} \&
  {Norman}}{1992a}]{stone1992a}
{Stone} J.~M.,  {Norman} M.~L.,  1992a, ApJS, 80, 753

\bibitem[\protect\citeauthoryear{{Stone} \& {Norman}}{{Stone} \&
  {Norman}}{1992b}]{stone1992b}
{Stone} J.~M.,  {Norman} M.~L.,  1992b, ApJS, 80, 791

\bibitem[\protect\citeauthoryear{{Strang}}{{Strang}}{1968}]{strang1968a}
{Strang} G.,  1968, SIAM J. Numer. Anal., 5, 506

\bibitem[\protect\citeauthoryear{{Tasker}, {Brunino}, {Mitchell}, {Michielsen},
  {Hopton}, {Pearce}, {Bryan} \& {Theuns}}{{Tasker} et~al.}{2008}]{tasker2008a}
{Tasker} E.~J.,  {Brunino} R.,  {Mitchell} N.~L.,  {Michielsen} D.,  {Hopton}
  S.,  {Pearce} F.~R.,  {Bryan} G.~L.,    {Theuns} T.,  2008, MNRAS, 390, 1267

\bibitem[\protect\citeauthoryear{{Teyssier}}{{Teyssier}}{2002}]{teyssier2002a}
{Teyssier} R.,  2002, Astron. and Astrophys., 385, 337

\bibitem[\protect\citeauthoryear{{Toro}}{{Toro}}{1997}]{toro1997a}
{Toro} E.,  1997, {Riemann solvers and numerical methods for fluid dynamics.}.
Springer

\bibitem[\protect\citeauthoryear{{Truelove}, {Klein}, {McKee}, {Holliman} II,
  {Howell}, {Greenough} \& {Woods}}{{Truelove} et~al.}{1998}]{truelove_etal98}
{Truelove} J.~K.,  {Klein} R.~I.,  {McKee} C.~F.,  {Holliman} II J.~H.,
  {Howell} L.~H.,  {Greenough} J.~A.,    {Woods} D.~T.,  1998, ApJ, 495, 821

\bibitem[\protect\citeauthoryear{{van Leer}}{{van Leer}}{1977}]{van_leer1977a}
{van Leer} B.,  1977, Journal of Computational Physics, 23, 276

\bibitem[\protect\citeauthoryear{{Vietri}, {Ferrara} \& {Miniati}}{{Vietri}
  et~al.}{1997}]{vietri1997a}
{Vietri} M.,  {Ferrara} A.,    {Miniati} F.,  1997, ApJ, 483, 262

\bibitem[\protect\citeauthoryear{{Wada} \& {Norman}}{{Wada} \&
  {Norman}}{1999}]{wada_norman99}
{Wada} K.,  {Norman} C.~A.,  1999, ApJL, 516, L13

\bibitem[\protect\citeauthoryear{{Wadsley}, {Veeravalli} \&
  {Couchman}}{{Wadsley} et~al.}{2008}]{wadsley2008a}
{Wadsley} J.~W.,  {Veeravalli} G.,    {Couchman} H.~M.~P.,  2008, MNRAS, 387,
  427

\bibitem[\protect\citeauthoryear{{Wang}, {Abel} \& {Zhang}}{{Wang}
  et~al.}{2008}]{wang2008a}
{Wang} P.,  {Abel} T.,    {Zhang} W.,  2008, ApJS, 176, 467

\bibitem[\protect\citeauthoryear{{Yang}}{{Yang}}{1990}]{yang1990a}
{Yang} H.,  1990, Journal of Computational Physics, 89, 125

\bibitem[\protect\citeauthoryear{{Yepes}, {Kates}, {Khokhlov} \&
  {Klypin}}{{Yepes} et~al.}{1997}]{yepes_etal97}
{Yepes} G.,  {Kates} R.,  {Khokhlov} A.,    {Klypin} A.,  1997, MNRAS, 284, 235

\end{thebibliography}

\label{lastpage}

\end{document}